\documentclass[a4paper,11pt]{article}
\pdfoutput=1 

\usepackage{jcappub} 

\usepackage[T1]{fontenc} 

\usepackage[T1]{fontenc}

\usepackage[english, english]{babel}

\DeclareRobustCommand{\VAN}[3]{#2}
\let\VANthebibliography\thebibliography
\def\thebibliography{\DeclareRobustCommand{\VAN}[3]{##3}\VANthebibliography}

\usepackage{algorithm} 
\usepackage[noend]{algpseudocode} 

\usepackage[bitstream-charter]{mathdesign}
\usepackage[usenames]{color}
\usepackage{xcolor}
\usepackage{soul}
\usepackage{multirow}
\usepackage{graphicx}
\usepackage{wrapfig}   %
\usepackage{amsmath}
\usepackage{nicefrac}
\usepackage{microtype}
\usepackage{amssymb}
\usepackage{physics}
\usepackage{bbding}
\usepackage{csquotes} 
\usepackage{float}
\usepackage{xurl}
\usepackage{adjustbox}
\usepackage{comment}
\usepackage{hyperref}
\usepackage{tensor} 
\usepackage{enumerate}
\usepackage{enumitem}
\usepackage{orcidlink} 
\usepackage{booktabs} 
\usepackage[font=small,labelfont=bf]{caption}
\usepackage{subcaption}
\usepackage{pdfpages}

\hypersetup{colorlinks=true, citecolor=blue, linkcolor=blue, urlcolor=blue}

\definecolor{hotpink}{RGB}{255, 105, 180}

\definecolor{orcidlogocol}{HTML}{A6CE39}

\DeclareSymbolFont{usualmathcal}{OMS}{cmsy}{m}{n}
\DeclareSymbolFontAlphabet{\mathcal}{usualmathcal}

\title{\boldmath Starobinsky in Stereo: \\SKA-CMB Synergy in SBI}

\author[a,1]{Benedikt Schosser\,\orcidlink{0009-0007-8905-7749},\note{Corresponding author.}}
\author[b]{Caroline Heneka\,\orcidlink{0000-0001-8883-0583},}
\author[a,c]{and Bj{\"o}rn Malte Sch{\"a}fer\,\orcidlink{0000-0002-9453-5772}}

\affiliation[a]{Zentrum f{\"u}r Astronomie der Universit{\"a}t Heidelberg, Astronomisches Rechen-Institut,\\Philosophenweg 12, 69120 Heidelberg, Germany}
\affiliation[b]{Institut für Theoretische Physik, Universit{\"a}t Heidelberg,\\Philosophenweg 16, 69120 Heidelberg, Germany}
\affiliation[c]{Interdisziplin{\"a}res Zentrum f{\"u}r wissenschaftliches Rechnen, Universit{\"a}t Heidelberg,\\INF205, 69120 Heidelberg, Germany}

\emailAdd{schosser@stud.uni-heidelberg.de}
\emailAdd{heneka@thphys.uni-heidelberg.de}
\emailAdd{bjoern.malte.schaefer@uni-heidelberg.de}

\abstract{
Modern machine learning techniques can unlock the vast cosmological information encoded in forthcoming Square Kilometre Array (SKA) observations. We show that tomographic 21\,cm data from the reionisation era can yield stringent tests of inflationary models -- here illustrated with Starobinsky $R+R^2$ inflation. Using a simulation-based inference (SBI) framework, we compare neural summaries (convolutional network and vision transformer) with a traditional power spectrum summary and perform a fully joint SBI analysis combining 21\,cm data with data of the cosmic microwave background (CMB). Forecasts based on realistic mock observations indicate that SKA alone will achieve constraints competitive with Planck, and that the combined SKA + CMB dataset will tighten bounds on both inflationary and $\Lambda\mathrm{CDM}$ parameters considerably while improving precision on key astrophysical quantities.
}

\begin{document}
\maketitle
\flushbottom

\section{Introduction}
\label{sect:intro}
The 21\,cm line, originating from the forbidden spin-flip transition of neutral hydrogen, is an exciting way to observe a large part of the observable Universe. It promises a window into Cosmic Dawn and the Epoch of Reionisation at $z\gtrsim 5$~\cite{Furlanetto:2006jb}. With the 3D-tomography from the upcoming Square Kilometre Array (SKA)\footnote{\url{https://www.skao.int/en}} the understanding of cosmic structure formation, galaxy evolution, the thermal history of the intergalactic medium, and many other processes, will be substantially improved. While its potential for astrophysics is already well explored, the power of SKA to test theories about fundamental cosmology has received less attention in comparison. 

Among those theories, cosmic inflation stands out. It elegantly solves the flatness and horizon problem, while seeding the large-scale structure~\cite{sato_inflation}. In the sea of inflationary models, Starobinsky inflation~\cite{Starobinsky:1980te, Starobinsky:1983zz} is a popular choice due to its relative simplicity and accurate predictions~\cite{Planck:2018_inflation}. The model extends general relativity by a quadratic term in the Ricci-scalar. 

This sets the stage for a typical Bayesian inference task, with the goal of finding the posterior through Bayes' theorem
\begin{align}
    p(\theta\mid y) = \frac{\mathcal{L}(y\mid  \theta)p(\theta)}{p(y)}\,,
\end{align}
with the likelihood $\mathcal{L}(y\mid \theta)$, prior $p(\theta)$ and the evidence $p(y)$ depending on parameters $\theta$ and data $y$. The complexity in modern cosmological probes, caused by non-Gaussianities in the data, systematics and foregrounds, often results in an intractable likelihood. A forward-model needs to take its role and one arrives at simulation-based inference (SBI)~\cite{papamakarios2018fastepsilonfreeinferencesimulation,Cranmer_2020}. 

Two ingredients are central to precision cosmology: highly informative summary statistics and the joint analysis of independent datasets. The non-Gaussian nature of the 21\,cm data requires summary methods more sophisticated than the power spectrum. Neural networks have proven to be able to capture ever more information and optimise their learned representation, or summary statistic, towards the goal of optimal, well-calibrated posteriors~\cite{neutsch:2022hmv,Schosser:2024aic,Ore:2024jim}. The cosmic microwave background (CMB) is perfectly summarised by the angular power spectrum and a likelihood exists~\cite{Planck:2019nip}, nevertheless, we are able to jointly analyse this legacy dataset within the SBI framework with 21\,cm mock data.\\
\\
\textbf{Our contributions}
\begin{itemize}[noitemsep, topsep=0pt]
    \item Realistic forecasted constraints on Starobinsky Inflation from SKA data
    \item Joint inference of two cosmological datasets with flow matching posterior estimation
    \item Information-theoretic assessment of summary statistics for 21\,cm tomography 
    \item Incorporate inflationary models in existing 21\,cm and Boltzmann codes\\
\end{itemize}
\textbf{Related Work}\quad
SBI is nowadays commonly used in cosmological analyses~\cite{Villaescusa:2022,Reza:2024djq, Zubeldia:2025qlt, vonWietersheim-Kramsta:2024cks} also related to 21\,cm data~\cite{Hothi:2023abe,Schosser:2024aic,Ore:2024jim, Prelogovic:2023uww,Saxena:2023tue,Zhao:2023uvf,Zhao:2023tep} to avoid potentially biased constraints from 21cm power spectrum MCMC analysis~\cite{Zhao:2021ddh}. Cosmological forecasts for the SKA have been studied in Refs.~\cite{Bull:2014rha,Liu:2019srd,Bacon2020SKA}. Further forecasts have been carried out for neutrino masses~\cite{Autieri:2025sxz}, modifications to general relativity~\cite{Heneka:2018kgn} and coupled quintessence~\cite{Liu:2019ygl}. The effect of 21\,cm data on CMB constraints due to reinisation information was studied in Ref.~\cite{Liu:2015txa}. An extension to Starobinsky inflation was constrained with a simplified SKA likelihood in Ref.~\cite{Modak:2022gol}, while the Starobinsky model gained again popularity in the light of recent ACT results~\cite{Drees:2025ngb,km3q-rm34}. Optimal summary statistics are researched in many cosmological sub-fields~\cite{Lehman:2024vyl,Lanzieri:2024mvn,Makinen:2024xph} with special focus on 21\,cm data~\cite{neutsch:2022hmv,Hothi:2023abe,Schosser:2024aic,Ore:2024jim,schiller:2025,Prelogovic:2024ips,Diao:2024wrf}. Neural networks have been used for Planck data~\cite{Wolz:2023gql} and most recently a method for joint SBI was proposed in Ref.~\cite{FrancoAbellan:2025fkb}, which still requires MCMC samples.\\
\\
The remainder of this paper is organised as follows: In Section~\ref{sect:staro_inflation} we recapitulate the Starobinsky model of inflation. Section~\ref{sect:npe} introduces neural posterior estimation in the conditional flow matching framework and explains joint inference in SBI. The simulation pipeline is described in Section~\ref{sect:data_sim}, before discussing our neural networks and training scheme in Section~\ref{sec:impl}. We validate the method against MCMC in Section~\ref{sec:res_cmb}. In Section~\ref{sec:res_mi} we find the optimal summary and show its calibration in Section~\ref{sec:val_cal}. Finally, in Section~\ref{sec:constraints} we display the forecasted constraints on Starobinsky inflation before discussing our findings in Section~\ref{sec:conclusion}.
\begin{figure*}[t!]
    \centering
    \includegraphics[page=1, width=\linewidth]{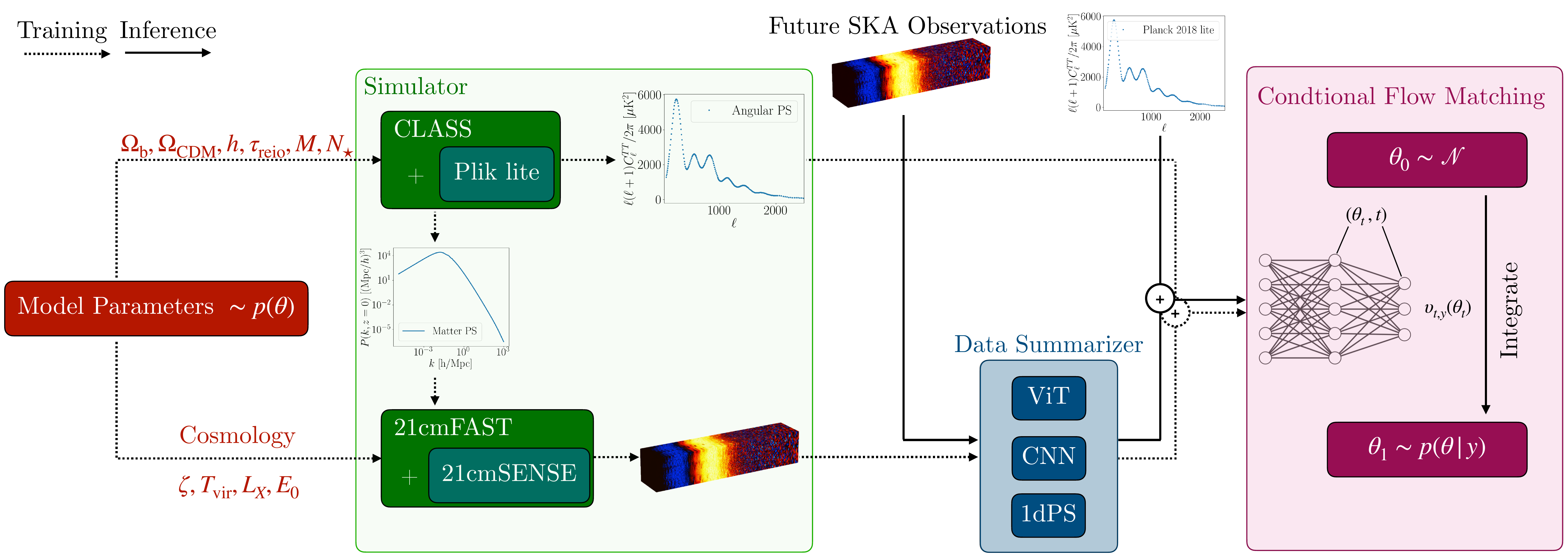}
    \caption{Inference pipeline to constrain Starobinsky Inflation with SKA and Planck. During training (dashed lines) model parameters are given to the simulator, a combination of CLASS and 21cmFAST, yielding a mock angular power spectrum (upper plot) and a 21\,cm light cone (lower plot). The light cone is summarised by one of three methods, before it is combined with the spectrum. A neural network is optimised to learn the vector field from a base distribution to the posterior. During inference (solid lines) the observations pass the same route as the simulations before samples from the base distribution are transformed into samples from the posterior.}
    \label{fig:overview}
\end{figure*}
\section{Starobinsky Inflation}
\label{sect:staro_inflation}
The Starobinsky model of inflation~\cite{Starobinsky:1980te, Starobinsky:1983zz, kallosh2025presentstatusinflationarycosmology} can explain a phase of exponential expansion in the early universe by a small modification of the action of general relativity. It is one of the simplest, yet currently best-fitting models for the  data~\cite{Planck:2018_inflation}. One new parameter is introduced to the cosmological model by modifying the action as
\begin{align}
    S_J = \frac{M_\mathrm{Pl}^2}{2}\int\dd^4x \sqrt{-g_J} \left( R + \frac{1}{6M^2}R^2 \right)\,,
\end{align}
where the subscript $J$ denotes the Jordan frame, $g_J$ is the determinant of the metric, using the mostly-plus convention (-,+,+,+), $M_\mathrm{Pl} = (8 \pi G)^{-1/2}$ is the Planck mass and $M$ is the free parameter of the theory. It can be interpreted as a mass. This theory falls into the category of $f(R)$ models, where $f(R)$ can be any function of the Ricci scalar $R$~\cite{DeFelice:2010aj}.

This modification is connected to a scalar-tensor theory by a change to the Einstein frame, recasting the effects of the $R^2$ term into a canonical scalar degree of freedom. A Legendre and a conformal transformation yield the well-known action of single-field inflation
\begin{align}
    S_E = \int \dd^4 x \sqrt{-g_E}\left[ \frac{M_\mathrm{Pl}^2}{2} R_E - \frac{1}{2}g^{\mu \nu}_E (\nabla_\mu \varphi \nabla_\nu \varphi) - V_E(\varphi) \right]\,,
    \label{eq:field_action}
\end{align}
with the scalar field $\varphi$ and its potential
\begin{align}
    V_E(\varphi) = \frac{3 M_\mathrm{Pl}^2 M^2}{4}\left[1 - \exp\left(-\sqrt{\frac{2}{3}} \frac{\varphi}{M_\mathrm{Pl}}\right) \right]^2\,.
\end{align}
This duality has the advantage that the formalism to study single-field inflation is applicable to this extension of gravity~\cite{Baumann:2009ds}. We split the field into a background and a perturbed part $\varphi(x^\mu) = \bar{\varphi} + \delta \varphi(x^\mu)$, for spacetime coordinates $x^\mu$. The background field equation can be written as
\begin{align}
    \ddot{\bar{\varphi}} + 3H \dot{\bar{\varphi}} + V_{E,\bar{\varphi}} = 0\,,
\end{align}
with the Hubble function $H$ and the conventions that $\dot{\bar{\varphi}} = \dv{t}\bar{\varphi}$, with cosmic time $t$, and $V_{E,\bar{\varphi}} = \dv{\bar{\varphi}} V_E$. The Hubble function fulfils the two equations
\begin{align}
    H^2 &= \frac{1}{3M^2_\mathrm{Pl}} \left[ \frac{1}{2}\dot{\bar{\varphi}} + V_E \right] \,, \\
    \dot{H} &= -\frac{1}{2M^2_\mathrm{Pl}} \dot{\bar{\varphi}}^2\,.
\end{align}
The evolution of the perturbations is described by the wave equation~\cite{Bassett:2005xm}
\begin{align}
    \delta\ddot{\varphi} + 3H\delta\dot{\varphi} + \left[ \frac{k^2}{a^2} + V_{E,\bar{\varphi}\bar{\varphi}} - \frac{1}{a^3}\dv{t}\left(\frac{a^3}{H}\bar{\varphi}^2 \right)\right] \delta\varphi = 0\,,
    \label{eq:perturbations}
\end{align}
and relates them to the gauge-invariant curvature perturbations 
\begin{align}
    \mathcal{R} = \frac{H}{\dot{\bar{\varphi}}} \delta\varphi\,.
\end{align}
Here, $a(t)$ is the scale factor and $k$ the wave number. The primordial scalar power spectrum is defined as
\begin{align}
    \langle \mathcal{R}(k)\mathcal{R}(k^\prime)\rangle = (2\pi)^3\delta_D(k+k^\prime)P_\mathcal{R}(k)\,,
    \label{eq:prim_ps}
\end{align}
where 
\begin{align}
    P_\mathcal{R}(k)=\left| \mathcal{R}\right|^2\,.
\end{align}
This primordial spectrum provides the initial condition for CMB anisotropies and large-scale-structure power spectra. The number of $e$-folds of inflation after the pivot mode $k_\star=0.05\,\text{Mpc}^{-1}$ crossed the horizon (at time $a_\star$) is defined as
\begin{align}
    N_\star = \int_{a_\star}^{a_\mathrm{end}} \dd \ln{a}\,.
\end{align}
To confront the model with observations we must specify how long inflation lasts. Consequently, $N_\star$ enters the analysis as an additional free parameter.

\section{Neural Posterior Estimation}
\label{sect:npe}
SBI~\cite{papamakarios2018fastepsilonfreeinferencesimulation,Cranmer_2020} enables approximate posterior estimation when the likelihood is not tractable. Still, it is not limited to this scenario as it can bring performance improvements in any inference task~\cite{PhysRevLett.130.171403}. Neural posterior estimation (NPE) uses a conditional generative network to approximate the posterior without assuming a likelihood shape. 

\subsection{Conditional Flow Matching}
We use conditional flow matching (CFM) to learn the posterior \cite{lipman2023flowmatchinggenerativemodeling, dax2023flowmatchingscalablesimulationbased}, a training scheme that interpolates between a base and target distribution via a path driven by an ordinary differential equation (ODE). Conditional flow matching is based on continuous normalising flows \cite{NEURIPS2018_69386f6b}, which map from a base to a target distribution, parameterised by time $t \in [0,1]$. The target is the posterior $p(\theta\mid y)\approx q_{t=1}(\theta\mid y)$, and the base distribution is typically a simple one, which does not depend on the data $q_{t=0}(\theta)$. The flow $\phi_{t,y}(\theta)$ is defined by a vector field in the parameter space $\upsilon_{t,y}(\theta)$ as
\begin{align}
    \dv{\phi_{t,y}(\theta)}{t} = \upsilon_{t,y}(\phi_{t,y}(\theta))\,.
    \label{eq:vector_field}
\end{align}
Samples $\theta_t$ along this path are obtained by integrating the ODE. The density of the approximate posterior is given by
\begin{align}
    q(\theta\mid y) = [\phi_{1,y}]_\star q_0(\theta) = q_0(\theta) \exp \left(-\int_0^1 \operatorname{div} \upsilon_{t,y}(\theta_t) \,\dd t \right)\,,
\end{align}
with the change-of-variable operator $[\cdot]_\star$ and using the continuity equation for the probability density of the flow. Unlike normalising flows, $\upsilon_{t,y}(\theta)$ can be approximated by any neural network; there are no architectural constraints as long as $\upsilon_{t,y(\theta)}:\mathbb{R}^{n+d+1}\rightarrow \mathbb{R}^n$, with data and parameter dimensions $d$ and $n$, respectively.

The training objective is to regress $\upsilon_{t,y}$ on a vector field $u_{t,y}$. In Ref.~\cite{lipman2023flowmatchinggenerativemodeling}, it is shown that $u_{t,y}$ can be constructed as a mixture of many simple paths, making it accessible. The so-called sample-conditional path is defined by a vector field $u_t(\theta\mid \tilde{\theta})$ and its corresponding distribution $p_t(\theta\mid \tilde{\theta})$. The loss function reads 
\begin{align}
L_\mathrm{SCFM} = \mathbb{E}_{t\sim\mathcal{U}[0,1], y\sim p(y), \tilde{\theta}\sim p(\theta\mid y), \theta_t \sim p_t(\theta_t\mid \tilde{\theta})}\bigl\lVert \upsilon_{t,y}(\theta_t)-u_t(\theta_t\mid \tilde{\theta})\bigr\rVert^2\,.    
\label{eq:loss_scfm}
\end{align}
One can choose from many different paths, here we follow \cite{dax2023flowmatchingscalablesimulationbased} and use Gaussians, defining
\begin{align}
    p_t(\theta\mid \tilde{\theta}) = \mathcal{N}(\theta\mid \mu_t(\tilde{\theta}), \sigma_t(\tilde{\theta}) I_n)\,, 
\end{align}
with the time-dependent mean and standard deviation given by the optimal transport path
\begin{align}
    \mu_t(\tilde{\theta})=t\tilde{\theta},\quad \sigma_t(\tilde{\theta}) = 1 - (1 - \sigma_\mathrm{min}) t\,.
\end{align}
The sample-conditional vector field then has the form
\begin{align}
    u_t(\theta\mid \tilde{\theta}) = \frac{\tilde{\theta} - (1 - \sigma_\mathrm{min})\theta}{1 - (1-\sigma_\mathrm{min})t}\,,
\end{align}
which is generated by the conditional flow
\begin{align}
    \phi_t(\theta\mid \tilde{\theta}) = (1-(1-\sigma_\mathrm{min})t)\theta + t \tilde{\theta}
\end{align}
using Eq.~\eqref{eq:vector_field}. Evaluating Eq.~\eqref{eq:loss_scfm} requires samples from the data distribution and the posterior, which are typically not accessible. As usual in SBI, we can use Bayes theorem and make the replacement $\mathbb{E}_{y\sim p(y), \theta \sim p(\theta\mid y)}\rightarrow \mathbb{E}_{\theta \sim p(\theta), y\sim \mathcal{L}(y\mid \theta)}$, which makes the samples accessible through a simulator. This yields the flow matching posterior loss \cite{dax2023flowmatchingscalablesimulationbased}
\begin{align}
    L_\mathrm{FMPE} = \mathbb{E}_{t\sim\mathcal{U}[0,1], \tilde{\theta}\sim p(\theta), y \sim \mathcal{L}(y\mid \tilde{\theta}), \theta_0\sim q_{t=0}(\theta)}\Bigl\lVert \upsilon_{t,y}(\phi_{t}(\theta_0\mid \tilde{\theta}))-\Bigl(\tilde{\theta} - ( 1-\sigma_\mathrm{min})\theta_0\Bigr)\Bigr\rVert^2\,.
\end{align}
If the data dimension is huge, a second network $s(y)$ can be introduced to map the data onto a lower-dimensional summary. The output of the summary network is then the input to the vector field $\upsilon_{t,y}$, and they are jointly trained.

\subsection{Joint multi-probe inference}
\label{sec:multip}
In addition to its ability to derive the posterior without assuming a likelihood shape, our inference framework offers the advantage of joint posterior inference from multiple probes that include learnt implicit covariances. Typically, $y$ comes from one type of observation, however, the derivation only requires $y$ to be sampled from a likelihood $\mathcal{L}(y\mid \theta)$. Suppose we want to jointly analyse the data of two different probes, $y_{\mathrm A}$ and $y_{\mathrm B}$. We partition the parameters into $\theta_{\mathrm{AB}}$ (shared by both probes), $\theta_{\mathrm A}$ (affecting only $y_{\mathrm A}$), and $\theta_{\mathrm B}$ (affecting only $y_{\mathrm B}$). If the probes are conditionally independent given $\theta$ the likelihood factorises
\begin{align}
    \mathcal{L}(y\mid \theta) = \mathcal{L}(y_\mathrm{A}\mid \theta_\mathrm{AB},\theta_\mathrm{A})\mathcal{L}(y_\mathrm{B}\mid \theta_\mathrm{AB},\theta_\mathrm{B})\,,
\end{align}
but this factorisation is not required in practice: the simulations themselves encode any cross-correlation between $y_{\mathrm A}$ and $y_{\mathrm B}$. Therefore, the cross-covariance between different probes does not need to be explicitly modelled. In other words, as long as every training sample is produced with the same random draw of $\theta$, data $y_\mathrm{A}$ and $y_\mathrm{B}$ are forward-modelled using the parameters $\theta$ within our simulator, in order to combine the experiments (here, the angular power spectrum is simulated for the CMB together with 21\,cm light cones). The network automatically uses the correct joint likelihood -- independent or not.

Our workflow therefore remains unchanged: sample $\theta\sim p(\theta)$, simulate the paired dataset $(y_{\mathrm A},y_{\mathrm B})\sim \mathcal{L}(y_{\mathrm A},y_{\mathrm B}\mid\theta)$, and train the CFM on these joint realisations. Probe‑specific parameters that are irrelevant for a given likelihood term (e.g. $\theta_{\mathrm B}$ in the first factor) integrate out during marginalisation, so the posterior for each probe can still be recovered from the same set of simulations. Finally, joint multi-probe inference with NPE offers the advantage to provide full posterior samples at inference (evaluation). In comparison, neural ratio estimation (NRE) learns the likelihood-to-evidence ratio and requires sampling with the learned ratio at inference to learn weights for posterior normalisation.

\section{Simulations}
\label{sect:data_sim}
In order to use the SBI framework fast and exact simulations are crucial. We use two different datasets to infer the parameters of Starobinsky inflation, as well as a minimalistic dark-energy background cosmology and necessary astrophysics - $(i)$ CMB measurements by Planck~\cite{Planck:2018nkj} and $(ii)$ forecasted SKA data. Properly simulating data for these two quite different sets requires the use of, and adjustments to, two numerical codes. The Boltzmann code CLASS\footnote{\url{https://github.com/lesgourg/class_public}}~\cite{Diego_Blas_2011} is well equipped to solve the perturbation equation \eqref{eq:perturbations} and produce the primordial power spectrum. We have made modifications to the code to incorporate the Starobinsky model, which allows us to solve for any physical and analytic inflaton potential. Secondly, the semi-numerical code 21cmFASTv3\footnote{\url{https://github.com/21cmfast/21cmFAST}}~\cite{Murray:2020trn} produces realisations of the 21\,cm temperature fluctuations. For accuracy of the 21cm signal we did not assume the so-called post-heating regime. We modified the code to incorporate different inflationary models by providing the matter power spectrum, which is used to sample the initial conditions. The simulation pipeline is illustrated in the green box of Figure~\ref{fig:overview}. 

We can divide our 9d parameter set into three categories: $(i)$ inflation, $(ii)$ cosmology and $(iii)$ astrophysics. All parameters and their definitions are listed in Table~\ref{tab:parameters}. The parameters describing inflation (see Section~\ref{sect:staro_inflation}) and the cosmology are given to CLASS, which calculates the primordial power spectrum and evolves it to the time of recombination, where we obtain the angular power spectrum of the CMB. Additionally, we get the matter power spectrum at $z=0$. 21cmFAST uses it to sample the initial conditions of the simulation and combined with the cosmological parameters, it can produce the 21\,cm temperature fluctuations. Simulating with CLASS also requires $\tau_\mathrm{reio}$, which is completely determined by the astrophysical parameters. Therefore, we calculate it from these and consider it to be a derived parameter.
\begin{table*}[t]
    \centering
    \begin{tabular}{llll}
        \toprule
        Category      & Symbol                        & Description                                                       & Prior                                                                            \\
        \midrule
        \multirow[t]{2}{*}{Inflation}
                      & $M$                            & Mass parameter in Starobinsky inflation                       & $\mathcal{U}[0.95,\,1.55]\,\times10^{-5}M_{\rm Planck}$               \\
                      & $N_\star$                     & Number of $e$-folds at the pivot scale                              & $\mathcal{U}[45,\,75]$                                                            \\
        \addlinespace
        \multirow[t]{3}{*}{Cosmology}
                      & $\Omega_{\mathrm{b}}$         & Baryon density parameter                                          & $\mathcal{N}[0.04936,\,0.00085]$                                                  \\
                      & $\Omega_{\mathrm{CDM}}$       & Cold dark matter density parameter                                & $\mathcal{U}[0.25,\,0.35]$                                                        \\
                      & $h$                           & Dimensionless Hubble parameter & $\mathcal{U}[0.65,\,0.75]$                                                         \\
        \addlinespace
        \multirow[t]{4}{*}{Astrophysics}
                      & $\zeta$                       & Ionisation efficiency                                             & $\mathcal{U}[10,\,200]$                                                           \\
                      & $T_{\mathrm{vir}}$            & Minimum virial temperature                                        & $\log\mathcal{U}[10^{4},\,10^{5}]\,\mathrm{K}$                                        \\
                      & $L_X$                         & Specific X-ray luminosity                                         & $\log\mathcal{U}[10^{39},\,10^{42}]\,\mathrm{erg\,s^{-1}\,M_\odot^{-1}\,yr}$          \\
                      & $E_0$                         & X-ray threshold energy for self-absorption                        & $\mathcal{U}[100,\,1200]\,\mathrm{eV}$                                            \\
        \bottomrule
    \end{tabular}
    \caption{Model parameters, their physical meaning, and prior distributions.}
    \label{tab:parameters}
\end{table*}

To generate the data we sample the parameters from the priors defined in Table~\ref{tab:parameters}. All have very wide flat priors, except $\Omega_\mathrm{b}$, which takes a Gaussian prior around the Planck 2018 value~\cite{Planck:2018vyg}. We adopt this choice because it is expected that 21\,cm data by itself has little constraining power for the baryon density~\cite{Mao:2008ug}. When we use the CMB data, we re-weigh the samples, such that it reflects a flat prior.

\begin{wrapfigure}[12]{r}{0.3\textwidth}  
  \vspace{-12pt}                         
  \centering
  \includegraphics[width=\linewidth]{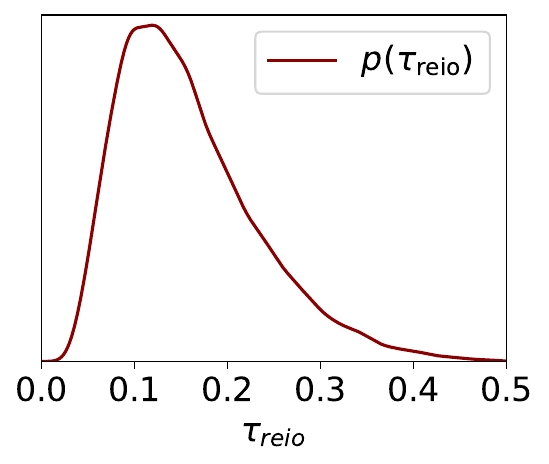} 
  \caption{Implicit prior on $\tau_\mathrm{reio}$ defined from the priors on the astrophysical parameters in Table~\ref{tab:parameters}}
  \label{fig:tau_reio}
  \vspace{-6pt}                    
\end{wrapfigure}
The CMB simulations have the same true labelled parameters as the 21\,cm simulations, i.e. they include the astrophysical parameters but not $\tau_\mathrm{reio}$. This comes with two implications for our analysis: we have an empirical prior on $\tau_\mathrm{reio}$, which is different from the typical Planck analysis, and we infer the astrophysical parameters from the Planck data and not $\tau_\mathrm{reio}$. Since there is no additional information in these parameters, because their only influence on the simulations is by providing the value of $\tau_\mathrm{reio}$, we can ignore them and marginalise over them. However, when we combine the datasets, we get the implicit information of the Planck data on the astrophysical parameter set and should see tighter constraints. Adding unnecessary parameters is harmless in a Bayesian analysis as the marginalisation process takes care of them.\\
\\
\textbf{CMB power spectrum data}\quad
Evolving the primordial power spectrum \eqref{eq:prim_ps} with a Boltzmann Code gives us the angular power spectrum at the time of recombination. We simulate up to $\ell=2508$. We add noise by sampling from a Gaussian with the covariance of the Plik lite high-$\ell$ TT likelihood~\cite{Planck:2019nip, Prince:2019hse}. The Plik lite likelihood averages the spectrum to 215 $\ell$ bins and marginalises the nuisance parameters. We only use the high-$\ell$ and TT part of the spectrum. \\
\\
\textbf{21\,cm light cone data}\quad
The 21\,cm light cones (LCs) are produced with 21cmFAST by evolving the initial density perturbations, given from the matter power spectrum from CLASS, and velocity conditions and evolves them with the Zel'dovich approximation. The result is a 3D LC of 21\,cm brightness temperature fluctuations $\delta T_b(x,y,z)$, with on-sky coordinates $x,y$ and the redshift $z$. We choose a box size of $200\,\mathrm{Mpc}$ and a cell size of $2.83\,\mathrm{Mpc}$, which results in 70 voxels in each spatial direction. The redshift is in the range of 5 - 35 with a resolution of 0.03 in equidistant steps. Therefore, the final LC has the shape (70, 70, 1000). We transform these simulations to mock observations with the code 21\,cmSense~\cite{21cmSense1, 21cmSense2}. The LC is produced by stitching coeval cubes together, these are transformed into Fourier space, where thermal noise is added. The noisy LC is then transformed back to real space. We consider the optimistic scenario where the foreground wedge in $k$-space covers only the instrument’s primary field-of-view and the thermal noise is based on 1080 hours of integrated SKA-Low stage 1 observations. To compare the analysis of the full light cone to power spectra, LCs are sliced in $\Delta z = 0.5$ steps, and the spherically-averaged power spectrum is calculated for 14 $k$-bins. This gives us 60 spectra.\\
\\
Our dataset consists of 32\,000 triplets, parameters, angular power spectrum and 21\,cm brightness fluctuations $\{ \theta,\,C_{\ell}^{TT},\, \delta T_b \}$, where noise is added to the simulated data to turn them into mock observations. An example of the $C_\ell^{TT}$ and $\delta T_b$ simulations can be seen in the upper and lower half of the simulator panel in Figure~\ref{fig:overview}, respectively.

\section{Implementation}
\label{sec:impl}

\textbf{Networks}\quad
This framework uses one or two neural networks, see Figure~\ref{fig:overview}. One network is used for the inference task and learns the vector field $\upsilon_{t,y}$, see Eq.~\eqref{eq:vector_field}. The other one finds an optimal summary of the high-dimensional data $s(y)$, which is not necessary for the spectra due to lower dimensionality. Conditional flow matching (Section~\ref{sect:npe}) is a highly efficient variant of SBI and only needs a simple network, we use a multi-layer perceptron (MLP) to approximate the vector field.

We test two different methods for $s(y)$. First, a 3D convolutional neural network (CNN) which has been extensively used in the context of 21\,cm light cones and is discussed in detail in Refs.~\cite{neutsch:2022hmv,Schosser:2024aic}. Second, a state-of-the-art Vision Transformer (ViT) introduced for 21\,cm data in Ref.~\cite{Ore:2024jim}. The exact architectures are listed in Appendix \ref{app:hyper}. The comparison allows us to quantify how much long-range attention (captured by ViTs) improves upon local convolutional features.\\
\\
\textbf{Pre-processing and data augmentations}\quad
Before training, we process our data to make it more efficient. All parameters are normalised to the unit interval. The 21\,cm temperature fluctuations are divided by 1250, which was found empirically in the data to have a smaller variance. Similarly, all CMB training spectra are used to compute a mean and standard deviation. These are then subtracted and divided out. Now, the common structure of the data is removed and the network can focus on the differences. We take a similar approach for the 1dPS. We rescale with the logarithm, then we find the minima and maxima in our training data, which are used to normalise the spectra to the unit interval. The input to our CFM has to be one-dimensional, therefore we flatten the $(60 \times 14)$ power spectra to $(840)$ entries. The learnt summaries are also one-dimensional, but much smaller. We test values in the range 9 ... 96, as discussed in Section~\ref{sec:res_mi}.

During training we use data augmentations. The 21\,cm light cones are randomly rotated and/or mirrored along the z-axis. We utilise the symmetry that the data possess to extract more information from the data we have. These geometric transforms leave the underlying cosmology invariant, effectively multiplying the size of the training set by a factor of eight. The noise for the CMB spectra is sampled at every iteration and added on top. We do not do this for the 21\,cm data as the noise calculation is more involved and computationally expensive. When we train for joint inference the output of the summary network and the normalised spectra are simply concatenated to one object and passed on to the CFM.\\
\\
\textbf{Training}\quad
The training scheme depends on whether a summary network is included. The summary network is required only when the raw input is prohibitively high-dimensional (e.g., 21\,cm light cones), for low-dimensional spectra the inference network alone suffices (see Figure~\ref{fig:overview}). If the 21\,cm data are summarised by the 1dPS or the CMB angular PS is used, we can immediately train the inference network in one go. When we want to summarise the data with a network, we find that a three-stage training scheme is most efficient. An overview of the training procedure and the hyperparameters is given in Table~\ref{tab:hyper_train}. First, we only train the summary network as a regression task. The goal is to minimise the mean squared error between the true labels of the simulation and the output of the network. This pre-training helps the network to learn the composition of the data while it only has to perform the simpler task of regression. In the second stage, we freeze the weights of the summary network but couple it to the inference network and optimise the FMPE loss, defined in Eq.~\eqref{eq:loss_scfm}. In this stage we do not need to load the data every epoch, which is the main bottleneck, and can efficiently optimise the inference network. During the regression phase, the output dimension is fixed to the number of parameters, but now we can choose any dimension. For the ViT we drop the MLP head, needed for regression, and the output dimension for the CNN is only adjusted in the next stage. 

Finally, both networks are trained together. Only then the correct convergence to Eq.~\eqref{eq:loss_scfm} can be guaranteed. Adjusting the size of summary-output dimension gives us the possibility to find an optimal embedding with the least information loss, see also Section~\ref{sec:res_mi}. Our staged strategy speeds convergence: stage 1 fits a good initial embedding, stage 2 optimises the posterior estimator while keeping I/O minimal, and stage 3 fine-tunes both components jointly.

For the joint inference of Planck and SKA data, we start with the networks trained on 21\,cm simulations. One could start from scratch if one is only interested in the joint inference, but since we already did the training for SKA data we can utilise this computational cost. With the summary network fixed we concatenate its output with the CMB spectra and train the inference network. We find that a simple concatenation is enough to include both datasets. Finally, we allow joint training and find the optimal representation. 

Our dataset of 32\,000 $\{ \theta,\,C_{\ell}^{TT},\, \delta T_b \}$ triplets is split into training, validation and test sets, consisting of 80\,\%, 15\,\% and 5\,\%, respectively. We optimise the networks with AdamW~\cite{loshchilov2019decoupledweightdecayregularization} and use the ones with minimal validation loss for testing. Training with full LCs was performed on one NVIDIA H100 GPU, while one NVIDIA A100 GPU was used for training with power spectrum data. For a detailed list of hyperparameters and training times see Table~\ref{tab:hyper_train}. 
 
\section{Results}
\label{sec:results}
To quantify how well the parameters of Starobinsky inflation will be constrained by the future SKA, and jointly with Planck, we first validate our SBI setup with MCMC and Planck data in Sect.~\ref{sec:res_cmb}. In Sect~\ref{sec:res_mi} the different summaries of 21\,cm data are studied from am information-theoretic point of view. The necessary validation and calibration of the SBI framework is shown in Sect~\ref{sec:val_cal} before we discuss our obtained posteriors in Sect.~\ref{sec:constraints}. Together, these subsections establish that
\begin{itemize}[noitemsep, topsep=0pt]
    \item our SBI implementation reproduces standard MCMC results for Planck data,
    \item neural summaries preserve more information than handcrafted statistics,
    \item the calibrated posteriors are reliable and can be trusted for forecasting,
    \item SKA will constrain cosmological and inflationary parameters to high precision,
    \item joint inference with Planck data reduces uncertainty in all parameters.
\end{itemize}

\subsection{Validation on Planck Data}
\label{sec:res_cmb}
\begin{figure}
    \centering
    \includegraphics[width=.6\textwidth]{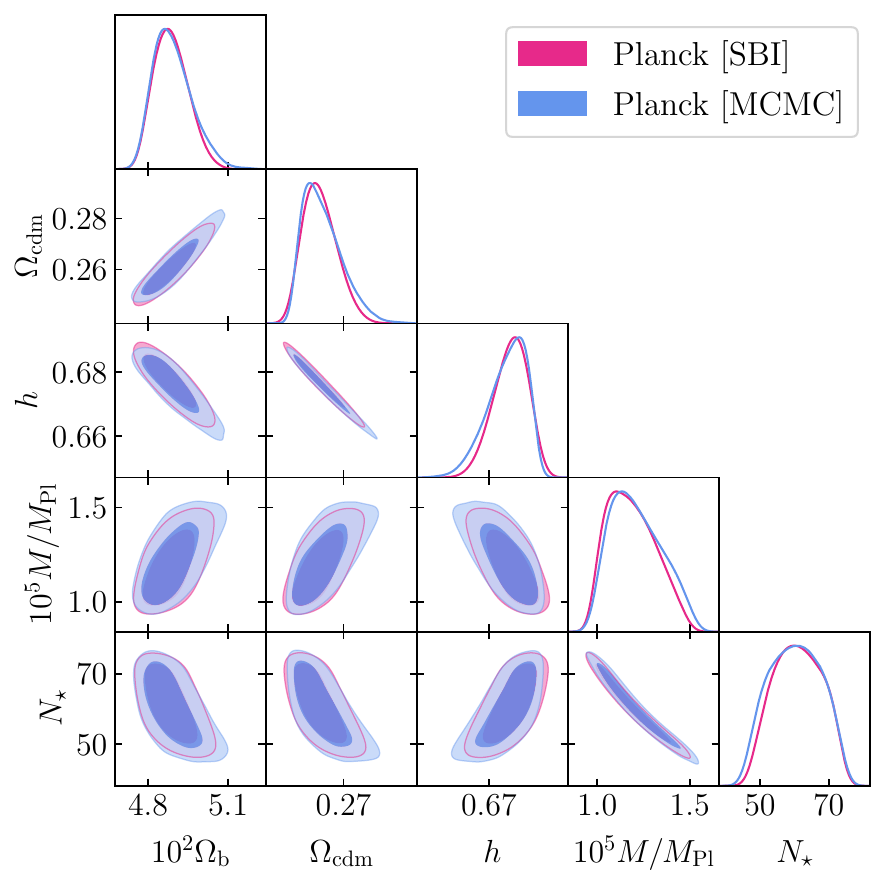}
    \caption{Comparison of SBI and MCMC for Planck data with the Plik lite TT high-$\ell$ likelihood. Both posteriors have $\approx$ 500\,000 samples. MCMC requires one CLASS call for each sample. SBI used 32\,000 parallelisable CLASS calls, eight minutes of training and 20 seconds for inference.}
    \label{fig:triangle_planck}
\end{figure}

First, we want to show how our SBI framework performs in a situation which can be described with traditional methods. The Planck inference pipeline does not need machine learning. It is possible to write down a likelihood and use any MCMC sampler to obtain the posterior, we use MontePython\footnote{\url{https://github.com/brinckmann/montepython_public}}~\cite{brinckmann2018montepython3boostedmcmc}. The prior on $\tau_\mathrm{reio}$ (see Figure~\ref{fig:tau_reio}) is handled with a Gaussian kernel density estimate on the samples. While a MCMC analysis is possible, it may be less efficient and more prone to tuning errors. The simulation pipeline, described in Section~\ref{sect:data_sim}, can be used to calculate the analytical likelihood, with the Plik covariance matrix, or get simulated datasets with Gaussian noise, from the same covariance matrix. The SBI method has the huge advantage that the data points do not depend on each other and can be simulated in parallel, while each MCMC sample depends on the previous. Additionally, the number of data samples is much fewer for the SBI method. Getting well converged chains can be a tricky task and we find that for this six-dimensional problem we need at least 100\,000 steps, of which only a small fraction are accepted. On the other side, the 32\,000 data points realised from parameter samples of the whole prior volume are plenty for our network to learn the posterior. Training time for our small inference network is $\approx$ 8 minutes, see Table~\ref{tab:hyper_train}. Therefore, we do not only save a huge amount of time and computational resources to obtain one posterior, but we can increase the samples to an almost arbitrary value and calculate the density. Additionally, we can statistically guarantee the correct posterior calibration and use any fiducial model in the prior volume. In Figure~\ref{fig:triangle_planck} we show the posteriors obtained from the MCMC run and the SBI method, which are almost identical. 

\subsection{Information Content Analysis}
\label{sec:res_mi}
To get the most constraining posterior, while being unbiased, it is necessary to use the most information in the data. Due to the large size of future 21\,cm observation, one needs to find an informative summary statistic. Mathematically, this is given by a sufficient statistic $t=T(y)$~\cite{fisher1922foundations}. The statistic $T$ is called sufficient for the parameters $\theta$ if the mutual information (MI) $I$ between the parameters and the data is equal to the MI between parameters and the summary statistic,
\begin{align}
    I(\theta,y) = I(\theta,T(y))\,,\quad\text{if $T$ is sufficient}\,,
\end{align}
where the mutual information for two continuous random variables is defined as
\begin{align}
    I(\theta,y) &= \int \int p(\theta,y) \log \frac{p(\theta,y)}{p(\theta) p(y)}\dd \theta \dd y.
    \label{eq:mi}
\end{align}
 In practice, we cannot evaluate this integral. However, we can evaluate it for the summary statistic. Rewriting Eq.~\eqref{eq:mi} with Bayes' theorem yields
 \begin{align}
     I(\theta,T(y)) &= \int\int p(\theta,T(y)) \log \frac{p(\theta\mid T(y))}{p(\theta)}\dd \theta \dd T(y) \\
     &= \mathbb{E}_{\theta,T(y)\sim p(\theta, T(y))}\left[ \frac{p(\theta\mid T(y))}{p(\theta)}\right]\,.
 \end{align}
We can evaluate the expectation value with Monte Carlo samples, while the posterior and prior density are accessible due to our SBI setup. Here, the statistic $T$ can be different neural networks (CNN or ViT), or handcrafted summaries like the power spectrum. The goal is to find networks such that $s(y)\approx T(y)$, where s(y) is the neural network. While we cannot claim that a statistic is sufficient, because $I(\theta,y$) is not accessible, we can find out, which has the most information about our parameter set. The obtained values, with Monte Carlo errors due to the finite sample size, are listed in Table~\ref{tab:mi_summary}.

A traditional summary statistic is the cylindrically-averaged power spectrum (1dPS). This choice mirrors many current 21 cm forecast studies and therefore serves as our benchmark. For each redshift slice, in equidistant steps of $\Delta z = 0.5$, we have 14 $k$-bins yielding a summary with $60\times14=840$ entries. 

The next summary statistic is learned by a 3D CNN and its dimension is chosen to be equal to the number of parameters (9). If there are no strong degeneracies between the parameters, an even lower-dimensional summary would be guaranteed to lose some information. We observe that increasing the dimension to 32 allows the network to pass more information to the CFM and the mutual information increases (+0.5 nats). Therefore, further information, e.g. noise, is captured by the additional dimensions. Using an even larger summary resulted in an over-confident network. The over-confidence manifests as SBC histograms that peak near 0 and 1, and an s-shaped TARP, indicating underestimated variance, see Appendix~\ref{app:val_method} for a detailed explanation. This could be prevented with more training data, however, it already indicates a saturation of information. Both CNN learned summaries bring a significant increase in MI over the 1dPS with an increase of $3$ and $3.5$ nats.

As the third method to summarise 21\,cm data, we use a ViT. Here, the dimension of the summary statistic is to a large degree determined by computational constraints. For this implementation we could choose between 48 and 96, where we observe the best performance for 48. Similar to the CNN, the setup becomes over-confident when the dimension is too large. The ViT manages to capture slightly more information than the CNN (+1.1 nats). This architecture is able to capture global features better than the CNN, which can help extracting all of the information. 

The binned angular PS of the CMB carries less information than the 1dPS of 21\,cm data, already indicating that the SKA measurements will be competitive with Planck, even across cosmological parameters. Combining the two datasets and learning a summary statistic of the 21\,cm data, which is optimised to compliment the CMB data, yields a huge increase in MI. It is important to note that MI is not additive. We observe the same performance differences between CNN and ViT as before.

This quantitative study of the information in summary statistics confirms that a power spectrum is not enough, even though it is high dimensional, to compress future SKA measurements and a learned one strongly outperforms it. The differences between the networks is small and is likely data-dependent.
\begin{table}[t]
  \centering
  \begin{tabular}{@{}lll@{}}  
    \toprule
    \textbf{Data} & \textbf{Summary Method  (Dimension)} & \textbf{Mutual Information [nats]} \\ \midrule
    SKA & 1d PS (60$\times$14) & 1$1.202 \pm 0.082$\\
    SKA & CNN (9) & $14.223 \pm 0.085$ \\
    SKA & CNN (32) & $14.712 \pm 0.077$ \\
    SKA & ViT (48) & $15.882 \pm 0.087$\\ 
    Planck & Angular PS (215)& $9.960 \pm 0.055$ \\
    SKA + Planck  & CNN (32) + Angular PS (215) & $21.285 \pm 0.072$ \\
    SKA + Planck & ViT (48) + Angular PS (215) & $21.995 \pm 0.073$ \\
    \bottomrule
  \end{tabular}
  \caption{Mutual information between parameters and summaries for different methods and data}
  \label{tab:mi_summary}
\end{table}
\subsection{Calibration and Validation}
\label{sec:val_cal}
\begin{figure}[t]
    \centering
    \includegraphics[width=\textwidth]{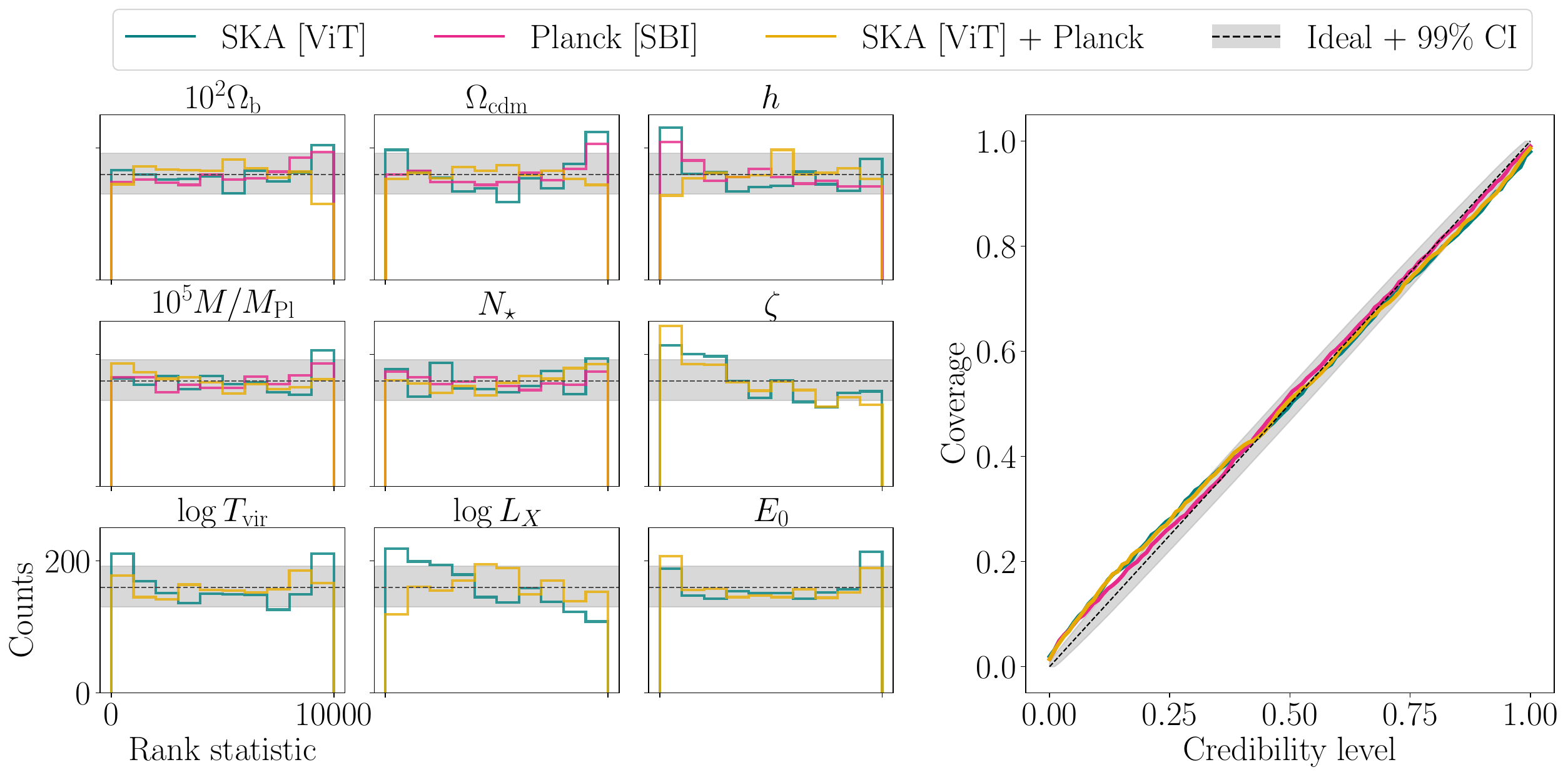}
    \caption{Validation metrics for different datasets, left SBC and right TARP. The black line indicates perfect calibration and the grey area the 99\,\% C.I. given the finite number of samples.}
    \label{fig:validation}
\end{figure}
\begin{figure}[t]
    \centering
    \includegraphics[width=\textwidth]{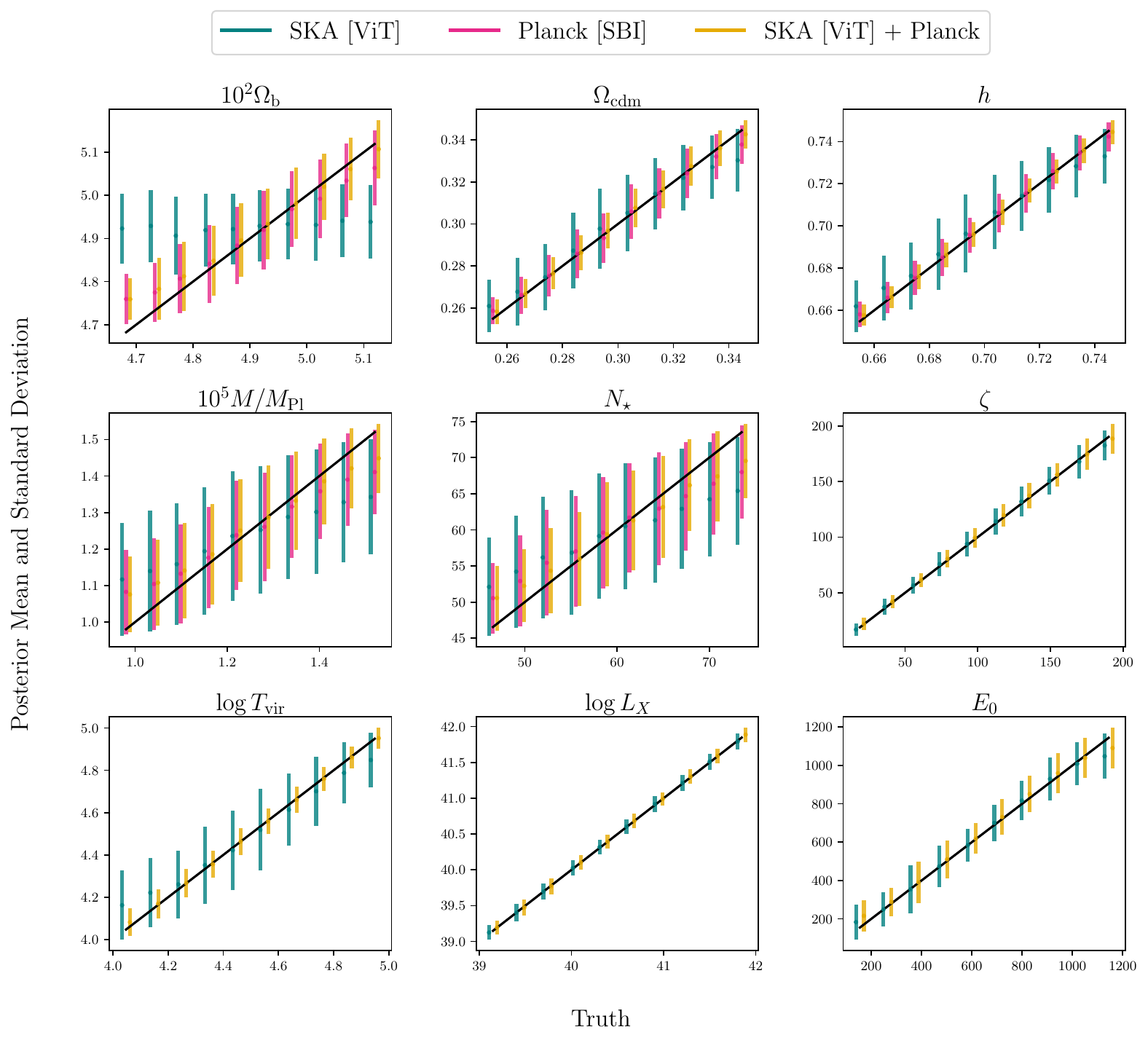}
    \caption{Parameter recovery for different datasets. The true parameter values (known from simulation) are plotted on the x-axis and the mariginalised 1d posteriors, illustrated as mean and standard deviation, are shown on the y-axis. The 1600 test data points are binned for better visualisation. Perfect recovery is indicated by the black line.}
    \label{fig:calibration}
\end{figure}
SBI has the great feature that its results can be statistically verified. We show the correct convergence of our networks using simulation-based calibration (SBC)~\cite{talts2020validatingbayesianinferencealgorithms} and the tests of accuracy with random points (TARP)~\cite{lemos2023samplingbasedaccuracytestingposterior}. While both methods use similar statistical arguments they have different failure modes and complement each other nicely. Together they test both marginal and joint coverage, closing common loopholes in posterior validation.

Both methods test for biases as well as under- and over-confidences in the learned posteriors. We use SBC to inspect the one dimensional marginalised and TARP for the full posterior. With SBC we get a clear insight which parameters might be biased and TARP confirms that we are not susceptible to marginalisation effects. The detailed algorithms are in Appendix~\ref{app:val_method}.

In Figure~\ref{fig:validation} (left) we show the SBC histograms for each of the nine parameters and the different datasets, using the ViT as summary network. A flat histogram corresponds to proper calibration. The grey 99\,\% C.I. error band is obtained from a binomial distribution due to the finite number of samples. The TARP curve is shown in Figure~\ref{fig:validation} (right), where a diagonal line shows the ideal coverage. The 99\,\% C.I. error band is again calculated from a binomial distribution. With these plots we can claim that we have properly calibrated networks and can trust the obtained posterior distributions. The validation plots for the different summary methods are listed in the Appendix in Figure~\ref{fig:val_methods}, which are all well calibrated.

Another performance check is parameter recovery or calibration. Each data point of the test set is put through the inference pipeline and the posterior is compared against the true parameter values. This is only possible in an amortized inference setup, as each posterior would require its own MCMC analysis. We show the true parameter values against the 1d marginalised posterior means with their standard deviation, obtained from 10\,000 samples. We have 1\,600 test samples, which are averaged into 10 bins for better visualisation. The error bars are calculated from the average standard deviation and its spread within the bin. In Figure~\ref{fig:calibration} we show the parameter recovery for the best-performing networks and the different datasets. The slight horizontal shift of points is purely for visualisation purposes. The black line shows perfect recovery. However, if all mean values would lie exactly on this line, we would have an incorrect coverage. Some spread is statistically expected and a sign of a well-calibrated network.

Here, we can already discuss the constraining power of SKA, Planck and their combination for the 1d distributions, while the full posterior for one fiducial model is discussed in Section~\ref{sec:constraints}. There is close to no information on $\Omega_\mathrm{b}$ in the 21\,cm data, as only the prior is recovered. All other parameters are well recovered, however, we see strong prior effects in the 1d distributions of $M$ and $N_\star$. They are strongly correlated and their individual boundaries are influenced by the prior choices, while their combination is tightly constrained. The addition of Planck data to the SKA inference reduces the uncertainty quite drastically, even for the astrophysical parameters. This is discussed in more detail in Section~\ref{sec:constraints}.

We show the influence of different summary methods on the parameter recovery in Figure~\ref{fig:cal_methods} in the Appendix. The neural networks clearly outperform the power spectrum summary. $\Omega_\mathrm{cdm}$, $h$ and $T_\mathrm{vir}$ are slightly better recovered by the CNN setup. It seems that the ViT is slightly more capable of learning correlations between the parameters, resulting in the higher mutual information, while the CNN recovers tighter 1d distributions. However, their difference is minor and with more training data or further hyperparameter optimisation their differences might decrease further.

\subsection{Forecasted Constraints}
\label{sec:constraints}
Finally, we can test how strong the constraints from SKA on Starobinsky inflation can be, given our assumptions on the noise model. As the fiducial model we choose the best-fit from our MCMC analysis of the Planck data. This leaves some freedom in the astrophysical parameters, as only their combination needs to recover a certain $\tau_\mathrm{reio}$. We use the following values:
\begin{align}
    & \Omega_\mathrm{b}^\mathrm{fid} = 0.0483\,,\quad
      \Omega_\mathrm{cdm}^\mathrm{fid} = 0.2538 \,,\quad
      h^\mathrm{fid} = 0.6819\,,\quad
      M^\mathrm{fid} = 1.0839\times 10^{-5}M_\mathrm{Pl} \,,\quad
      N_\star^\mathrm{fid} = 67.2020\,,\notag\\
    & \zeta^\mathrm{fid} = 118\,,\quad
      \log T_\mathrm{vir}^\mathrm{fid} = 4.2\,\mathrm{K},\quad
      \log L_X = 40\,\mathrm{erg\,s^{-1}\,M_\odot^{-1}\,yr}\,,\quad
      E_0 = 500\,\mathrm{eV} \,.
    \label{eq:fiducial}
\end{align}
\begin{figure}[t]
    \centering
    \includegraphics[width=\textwidth]{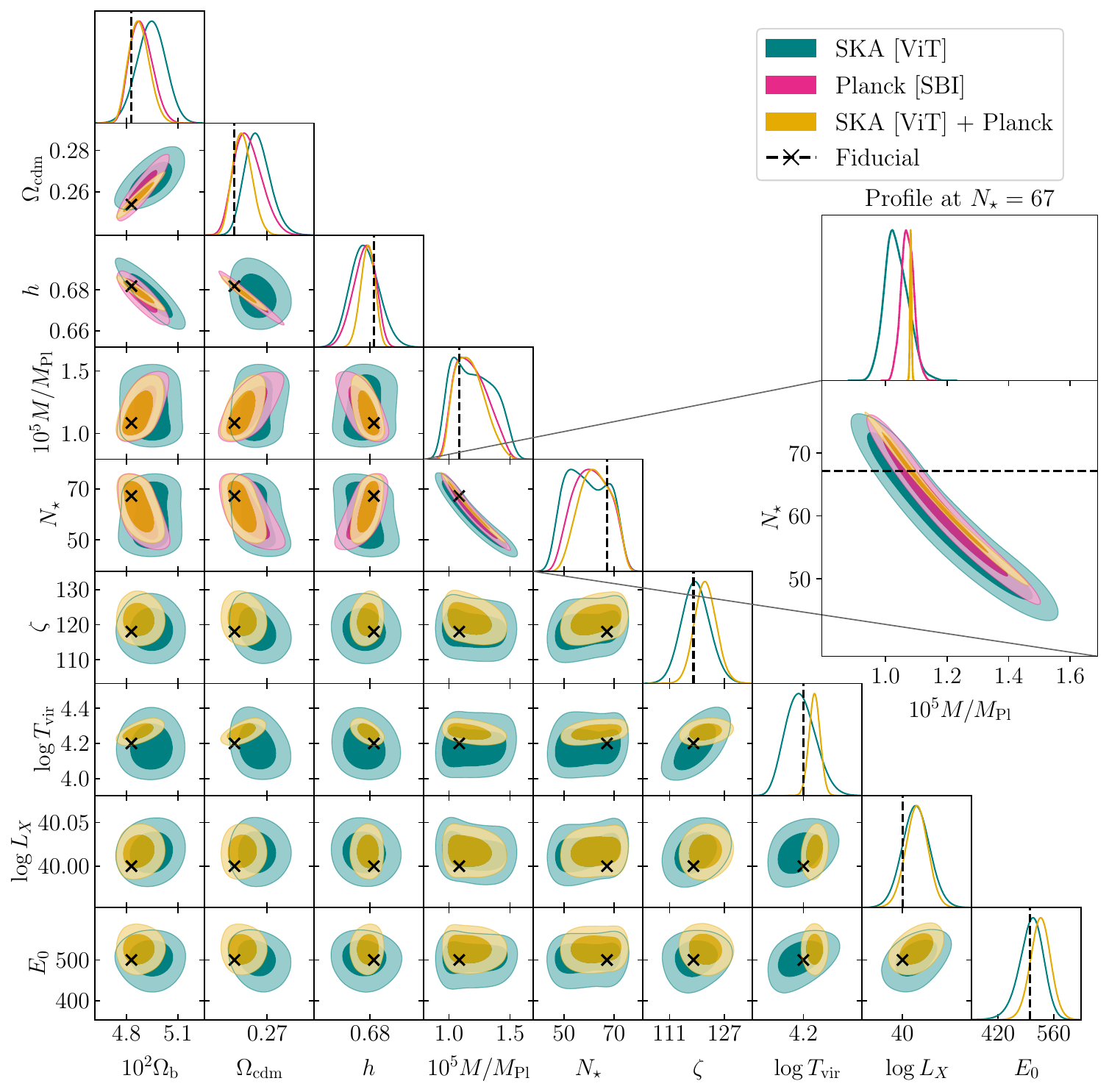}
    \caption{Posterior for cosmological and astrophysical parameters derived from mock SKA data (teal) using SBI, Planck data with SBI (pink) and joint SBI (yellow). The zoomed-in panel shows the the inferred correlation of $N_\star$ and $M$. We show the profiled distribution for a fixed $N_\star=67$ (the fiducial value) on top.}
    \label{fig:triangle_comp}
\end{figure}
Note that these values are slightly different from previous analyses, which is a result of the atypical prior on $\tau_\mathrm{reio}$. The derived posteriors are shown in Figure~\ref{fig:triangle_comp}. We have constraints derived from SKA (teal) and Planck (pink) alone as well as their combination (yellow). The Planck curves are only displayed for the cosmological and inflationary parameters, as the information on the remaining astrophysical ones is only implicit through $\tau_\mathrm{reio}$. Note that the posterior coming only from SKA has a Gaussian prior on $\Omega_\mathrm{b}$, while the samples coming from other experiments, here Planck CMB data and the joint analysis, are re-weighted to a flat prior.

First, we note that the constraints on the astrophysical parameters derived from the SKA mocks are comparable in size to those obtained in previous studies for smaller parameter sets -- using the same astrophysical parameters but including only $\Omega_\mathrm{m}$ and $m_\mathrm{WDM}$ for the cosmology -- while the correlations become weaker~\cite{Schosser:2024aic}. This is to be expected, when more degrees of freedom are included. 

Second, the SKA posterior is competitive with the Planck one, even though it is higher-dimensional in this analysis. In particular cosmological constraints on the Hubble parameter $h$ and dark matter density parameter $\Omega_\mathrm{cdm}$ are of similar size for both experiments, where for both experiments we assumed flat priors. However, the correlation between $h$ and $\Omega_\mathrm{cdm}$ is less pronounced and the posterior for $\Omega_\mathrm{b}$ is prior-dominated for SKA alone. 

The two inflationary parameters of this analysis, the mass parameter $M$ in Starobinsky inflation and the number of $e$-folds at the pivot scale $N_\star$, are slightly better inferred by the Planck experiment. However, their marginalisations are largely prior-dominated, as can be best seen in the box-like contour for SKA and the very wide distribution for Planck. This is due to their very strong correlation, which is visible in the slow-roll formalism where  $A_s \propto M^2 N_\star^2$. The scalar amplitude $A_s$ fixes their ratio and they become inversely proportional. Breaking this degeneracy would require more information on the reheating process. Given the common approach to do the analysis at a fixed $N_\star$ we also show the profiled distribution for $M$ at the fiducial value of $N_\star$ above the zoomed-in panel of Figure~\ref{fig:triangle_comp}; we note, that in this case both Planck and SKA are able to put strong constraints on $M$.

Combining the two datasets, Planck and SKA, brings large improvements in all parameters. It is noteworthy that the combination with Planck data can significantly reduce the uncertainty in all astrophysical parameters that describe the Epoch of Reionization and Cosmic Dawn, especially in the virial temperature $T_\mathrm{vir}$. While we only used the TT-spectra, one would expect that the polarisation information in the EE-spectra would reduce this even further. We also see a large improvement in the constraints of $h$ and $\Omega_\mathrm{cdm}$, where the marginalised one dimensional 95\,\% C.I. are reduced by a factor of 1.5 compared to Planck alone. Even $\Omega_\mathrm{b}$ becomes slightly better. 

Finally, the correlation contour between inflationary parameters $M$ and $N_\star$ is further narrowed for the combination of Planck with SKA data, and we get a 95\,\% highest-posterior-density interval conditioned on the fiducial $N_\star$ of
\begin{align*}
     M \in [1.0772,\,1.0871] \times 10^{-5}M_\mathrm{Pl}\quad \mid  \quad N_\star=67 \, \quad \text{SKA + Planck}\,,
\end{align*}
reducing the size of the 95\,\% credibility interval by a factor of 9.15 compared to Planck alone in this forecast.

\section{Conclusion}
\label{sec:conclusion}

The upcoming 21\,cm observations with the SKA promise an exciting new opportunity to test theories of gravity and inflation in a new way and at high redshifts. The sheer amount of recorded data will tighten parameter bounds, provided that we find sophisticated analysis methods to capture the full information content. 

In this study, we investigated the constraining power of the SKA on extended cosmological models, in particular the Starobinsky model of inflation, jointly with astrophysics of the Epoch of Reionization and Cosmic Dawn. We did so both alone and in combination with Planck CMB data. To perform this inference task, we modified existing Boltzmann and 21\,cm simulation codes and generated CMB angular power spectra and 21\,cm light cone data for the Starobinsky inflationary model. The intractable 21\,cm likelihood was addressed using simulation-based inference with conditional flow matching, which we have shown to achieve accurate and well-calibrated posterior estimation while being computationally efficient even when the likelihood is known.

We used mutual information to find the best summary statistic for a light cone and found that neural network summaries outperform power spectra. The two architectures considered, CNN and ViT, yielded very similar results, with a slight advantage for the ViT. Clearly, Gaussian statistics are insufficient to capture all information content and exploit SKA's full constraining power. 
Future work could also adopt a large range of noise models and employ the full Plik likelihood, incorporating the EE and TE spectra. 

Our forecasts show that SKA will be competitive with the results from the Planck collaboration, providing measurements at a very different range of redshifts, and despite the inclusion of additional astrophysical parameters. For example, SKA alone constrains $h$ and $\Omega_\mathrm{cdm}$ with precision comparable to Planck in our extended scenario. Combining the two datasets will increase the precision on all $\Lambda\mathrm{CDM}$ and inflationary parameters. For the inflationary parameters $M$ and $N_\star$ the $M-N_\star$ correlation contour is significantly narrowed. Profile constraints conditioned on $N_\star$ therefore significantly reduce the uncertainty on the mass parameter $M$ by a factor of $\mathcal{O}(10)$ compared to Planck alone. We also found that CMB data reduces the uncertainty on astrophysical parameters through its information on the reionisation history. 

Ultimately, the proposed approach yields fast and precise posteriors. Inference for one dataset is completed in a few seconds, whereas MCMC becomes unfeasible for 21\,cm constraints in extended parameter scenarios. Our SBI approach naturally allows the combination of multiple datasets and simplifies joint multi-probe analyses. Here, the SKA alone and in combination with CMB data was shown to provide precise constraints on both fundamental cosmological parameters and astrophysical processes at high redshifts.

\acknowledgments
We want to thank Sofia Palacios Schweitzer and Nathan Hütsch for valuable input on conditional flow matching, as well as, Ayodele Ore for help with the Vision Transformer.
\paragraph{Funding information}
This work was supported by the Deutsche Forschungsgemeinschaft (DFG, German Research Foundation) under Germany's Excellence Strategy EXC 2181/1 - 390900948 (the Heidelberg STRUCTURES Excellence Cluster) and by the Vector-Stiftung. CH’s work is funded by the Volkswagen Foundation. We acknowledge the usage of the AI-clusters {\em Tom} and {\em Jerry} funded by the Field of Focus 2 of Heidelberg University. 

\paragraph{Data availability}
The code will be publicly available on GitHub once the refereeing process is finished. The training data is shared upon request.

\appendix
\section{Additional Results}
\begin{figure}[h]
    \centering
    \includegraphics[width=\textwidth]{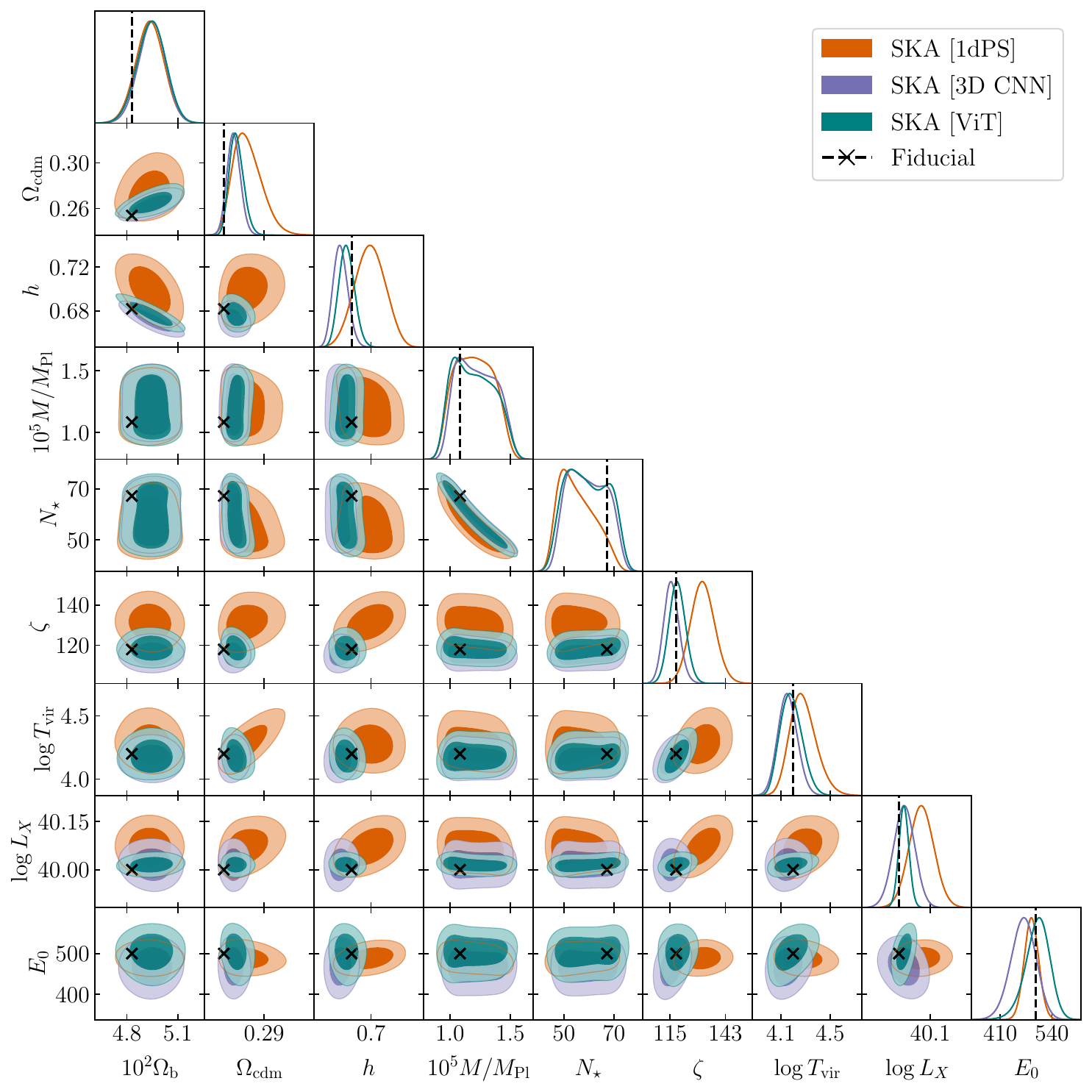}
    \caption{Posterior for cosmological and astrophysical parameters derived from mock SKA data. The 21\,cm light cone is summarised with the 1dPS (orange), a 3D CNN (purple) or a ViT (teal). }
    \label{fig:triangle_methods}
\end{figure}

\begin{figure}[h]
    \centering
    \includegraphics[width=\textwidth]{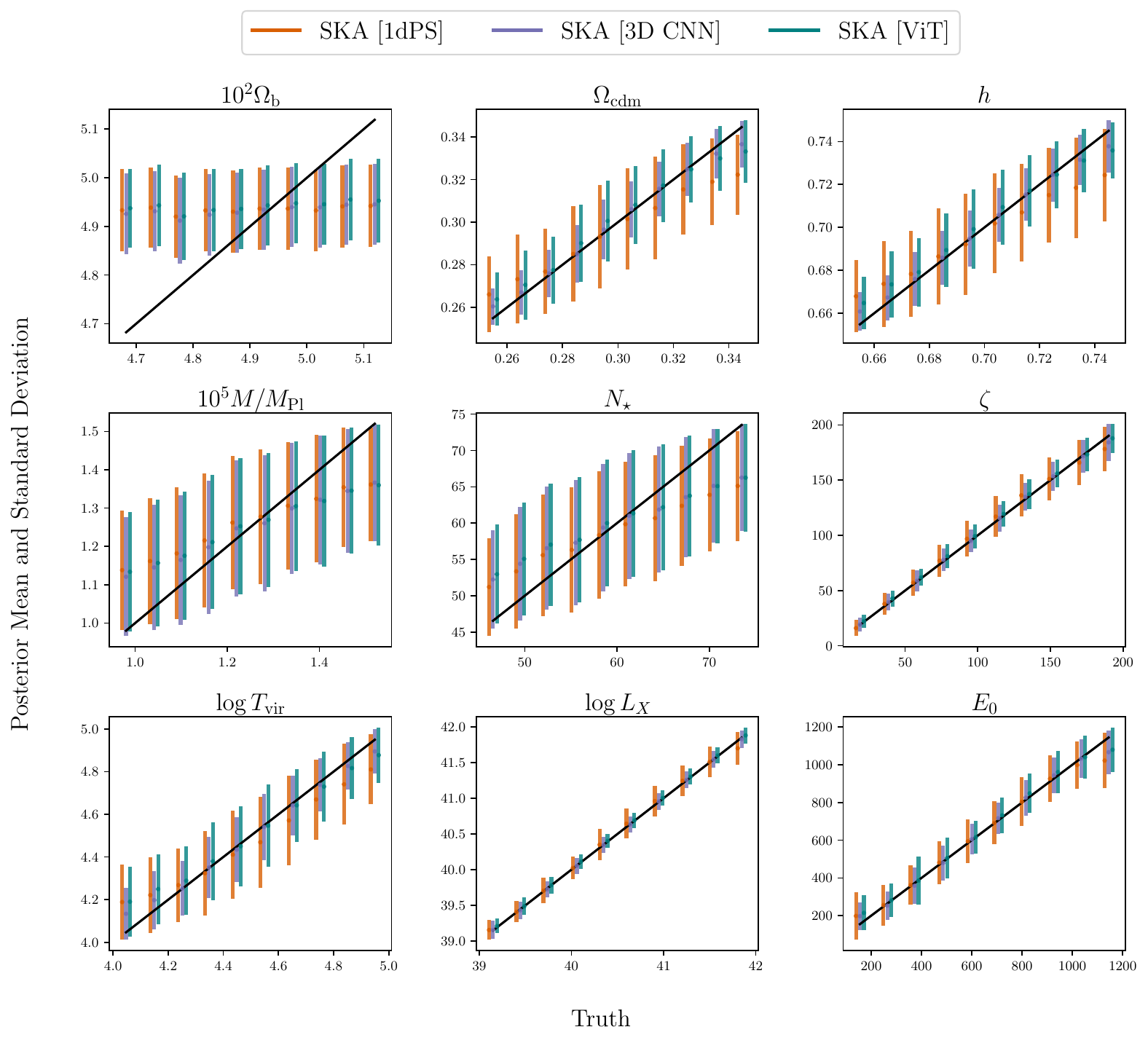}
    \caption{Parameter recovery from different summary networks for the SKA data. The true parameter values (known from simulation) are plotted on the x-axis and the mariginalised 1d posteriors, illustrated as mean and standard deviation, are shown on the y-axis. The 1600 test data points are binned for better visualisation. Perfect recovery is indicated by the black line.}
    \label{fig:cal_methods}
\end{figure}

\begin{figure}[h]
    \centering
    \includegraphics[width=\textwidth]{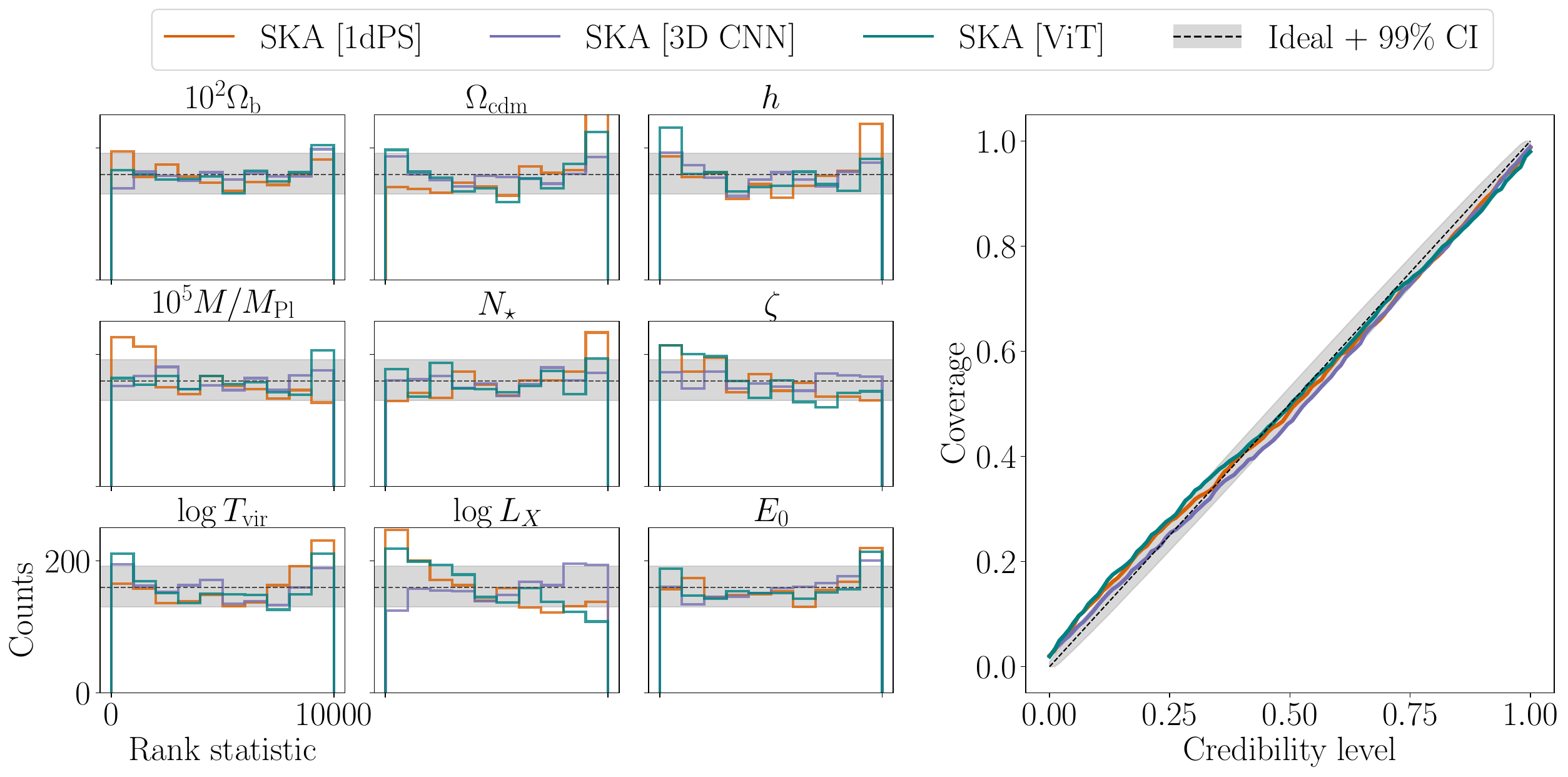}
    \caption{Validation metrics for different summary methods, left SBC and right TARP. The black line indicates perfect calibration and the grey area the 99\,\% C.I. given the finite number of samples.}
    \label{fig:val_methods}
\end{figure}

In order to determine the best summary statistic, based on mutual information, we trained and optimised the inference pipeline for each method. Here, we show the necessary calibration and validation plots (Figures~\ref{fig:cal_methods} and \ref{fig:val_methods}) as well as the full posterior (Figure~\ref{fig:triangle_methods}) for the fiducial model defined in Eq.~\eqref{eq:fiducial}. The two neural network summaries result in much tighter constraints, especially on the cosmological parameters. Interestingly, the power spectrum is able to capture the information on $E_0$ slightly better, for this fiducial model. The two networks yield very similar results, they only differ marginally for $L_X$ and $E_0$.
\label{app:additional_res}

\section{Validation Methods}
\label{app:val_method}
The validation methods employed in Section~\ref{sec:val_cal} are discussed here. We use simulation-based calibration (SBC)~\cite{talts2020validatingbayesianinferencealgorithms} and the tests of accuracy with random points (TARP)~\cite{lemos2023samplingbasedaccuracytestingposterior}. SBC diagnoses each 1d marginal, whereas TARP probes the full joint posterior. We prefer TARP to a conventional coverage test because it requires only posterior samples. Evaluating exact probabilities is possible, but for a CFM network it is far slower than for a normalising flow. Both diagnostics admit visual interpretation. In SBC, a sloped histogram indicates bias, a "$\cup$"-shape signals over-confidence, and a "$\cap$" under-confidence. In TARP, an S-curve about the diagonal denotes over-confidence, its mirror image under-confidence, and a systematic offset above or below the diagonal reflects bias. Pseudocode for SBC and TARP is provided in Algorithms \ref{alg:sbc} and \ref{alg:tarp}, respectively. The error bars are obtained from a binomial distribution. Each sample can be either in or out of the $\alpha$-credible region. Therefore, the expected variance due to the finite sample size $n$ can be modelled with the variance of a binomial distribution $B(n,\alpha)$.
\begin{algorithm}[H]
\caption{Simulation-Based Calibration}
\label{alg:sbc}
\begin{algorithmic}[1]
\Require Posterior sampler $p(\theta\mid y)$; test set $\{(y_i,\theta^\star_i)\}_{i=1}^{N}$; sample size $S$
\For{$i = 1,\dots,N$}
    \State Draw $\theta^{(1)},\dots,\theta^{(S)} \sim p(\theta\mid y_i)$
    \For{parameter index $j = 1,\dots,D$}
        \State \textbf{Rank statistic:}\;\;
              $r_{i,j} \leftarrow \sum_{s=1}^{S}\,
              \mathbb{I}\!\bigl[\theta^{(s)}_{j} < \theta^{\star}_{i,j}\bigr]$
        
    \EndFor
\EndFor
\State \Return\; $\{r_{i,j}\}$
\end{algorithmic}
\end{algorithm}

\begin{algorithm}[H]
\caption{TARP Coverage Evaluation}
\label{alg:tarp}
\begin{algorithmic}[1]
\Require Posterior sampler $p(\theta\mid y)$; test set $\{(y_i,\theta^\star_i)\}_{i=1}^{N}$;
         threshold grid $\{\alpha_k\}_{k=1}^{K}$; sample size $S$
\For{$i = 1,\dots,N$}
    \State Draw $\theta^{(1)},\dots,\theta^{(S)} \sim p(\theta\mid y_i)$
    \State Draw reference $\theta_{r}\sim p(\theta)$
    \State Radius $d \leftarrow \lVert\theta^{\star}_{i}-\theta_{r}\rVert$
    \State $f_i \leftarrow \frac{1}{S}\sum_{s=1}^{S}\mathbb{I}\!\bigl[\lVert\theta^{(s)}-\theta_{r}\rVert \le d\bigr]$
    \For{$k = 1,\dots,K$}
        \State $I_{i,k}\leftarrow\mathbb{I}\!\bigl[1-f_i\le\alpha_k\bigr]$
    \EndFor
\EndFor
\State \textbf{TARP coverage:}\;\;
      $\tau_k \leftarrow \frac{1}{N}\sum_{i=1}^{N} I_{i,k}, \quad k=1,\dots,K$
\State \Return\; $\{\tau_k\}_{k=1}^{K}$
\end{algorithmic}
\end{algorithm}
\clearpage
\section{Hyperparameters}
\label{app:hyper}

\begin{table}[h]
\vspace{-5pt}
\centering
\small
\begin{tabular}{l | l}
\toprule
\textbf{Parameter} & \textbf{Value} \\
\midrule
Input shape                             & $(70,\;70,\;1000)$ \\
Base channels $c$                       & 32 \\
Conv channels                           & $32\!\rightarrow\!32\!\rightarrow\!64\!\rightarrow\!64\!\rightarrow\!128\!\rightarrow\!128$ \\
Kernel sizes                           & $k_1=(2,2,43),\; k_2=(2,2,2)$ \\
Stride (first conv)                    & $(1,1,43)$ \\
Pooling                                 & MaxPool3D$(2,2,1)$ twice; AdaptiveAvgPool3D$(1,1,1)$ \\
Fully connected layers                  & $128 \rightarrow 128 \rightarrow 128 \rightarrow$ Output dimension \\
Output dimension         & stage dependent (see training table \ref{tab:hyper_train}) \\
\# Parameters &  310\,601 (9d) / 313\,568 (32d)\\
\bottomrule
\end{tabular}
\caption{Architecture of the \textbf{3D CNN} summary network}
\label{tab:arch_cnn}
\end{table}

\begin{table}[h]
\centering
\small
\begin{tabular}{l | l}
\toprule
\textbf{Parameter} & \textbf{Value} \\
\midrule
Input shape                         & $(70,\,70,\,1000)$ \\
Patch shape                         & $(5,\,5,\,50)$ \\
Number of patches                   & $14 \times 14 \times 20 = 3920$ \\
Embedding dimension                 & 48 \\
MLP hidden dim                      & 96 \\
Transformer depth (blocks)          & 4 \\
Attention heads per block           & 4 \\
Positional encoding                 & Learnable sin/cos \\
Patch aggregation                   & Mean \\
Summary dimension                   & 9 \\
Use extra head                      & stage dependent (see training table \ref{tab:hyper_train}) \\
(Head MLP layers, if used)          & $[48,\,48,\,9]$ \\
\# Parameters                       & 135\,240 (no head) / 138\,033 (with head) \\
\bottomrule 
\end{tabular}
\caption{Architecture of the \textbf{ViT} summary network.}
\label{tab:arch_vit}
\end{table}

\begin{table}[h]
\centering
\small
\begin{tabular}{l | l}
\toprule
\textbf{Parameter} & \textbf{Value} \\
\midrule
Hidden blocks                       & 6 $\times$ CondGLUMLP (residual) \\
Hidden dimension                    & 256 (per block) \\
Activation / gating                 & GLU \\
Output dimension & 9 \\
Input to first block                & $9 + \text{summary\_dim} + 1$ (time $t$) \\
ODE solver                          & \texttt{odeint} (rtol=1e-5, atol=1e-5) \\
Jacobian trace                      & brute-force autograd trace \\
\# Parameters (summary\_dim)         & 948\,352 (32) \\
                                    & 1\,024\,704 (48) \\
                                    & 1\,882\,750 (215) \\
                                    & 2\,059\,902 (247) \\
                                    & 2\,150\,014 (263) \\
                                    & 6\,084\,000 (840) \\
\bottomrule
\end{tabular}
\caption{Architecture of the \textbf{CFM}}
\label{tab:arch_cfm}
\end{table}
\begin{table}[ht]
  \centering\small
  \begin{tabular}{l||cc|cc|cc}
    \toprule
    \textbf{Stage} & \multicolumn{2}{c|}{\textbf{1}}
                   & \multicolumn{2}{c|}{\textbf{2}}
                   & \multicolumn{2}{c}{\textbf{3}} \\ 
    \midrule
    \textbf{Summary}    & ViT
                        & CNN 
                        & ViT
                        & CNN
                        & ViT
                        & CNN
                        \\ 
    \midrule
    Loss    & \multicolumn{2}{c|}{MSE}  
            & \multicolumn{2}{c|}{FMPE} 
            & \multicolumn{2}{c}{FMPE} 
            \\
    Trainable network   & ViT
                        & CNN 
                        & \multicolumn{2}{c|}{CFM} 
                        & ViT+CFM 
                        & CNN+CFM
                        \\
    Epochs  & 287
            & 983
            & 401
            & 970
            & 143
            & 472
            \\
    Learning rate $\times10^{-4}$   & 5
                                    & 10
                                    & 5
                                    & 5
                                    & 5
                                    & 5
                                    \\
    Scheduler   & \multicolumn{2}{c|}{Cosine Annealing}
                & \multicolumn{2}{c|}{Constant}
                & \multicolumn{2}{c}{Cosine Annealing}
                \\
    Batch size  & 256 
                & 1024
                & 256
                & 1024
                & 256
                & 1024
                \\
    Summary dimension   & \multicolumn{2}{c|}{9} 
                        & 48
                        & 9
                        & 48
                        & 32
                        \\
    Data    & \multicolumn{2}{c|}{SKA} 
            & \multicolumn{2}{c|}{SKA} 
            & \multicolumn{2}{c}{SKA} 
            \\
    Training time  & 19h\,01m 
                & 60h\,08m 
                & 10m 
                & 17m 
                & 11h\,40m 
                & 32h\,20m 
                \\ 
    \midrule
    \midrule
    \textbf{Stage} & \multicolumn{2}{c|}{\textbf{(4)}}
                   & \multicolumn{2}{c|}{\textbf{(5)}}
                   & \multicolumn{2}{c}{\phantom{xx}} \\ 
    \midrule
    \textbf{Summary}    & ViT
                        & CNN
                        & ViT
                        & CNN
                        & 1dPS
                        & Planck
                        \\ 
    \midrule
    Loss    & \multicolumn{2}{c|}{FMPE} 
            & \multicolumn{2}{c|}{FMPE} 
            & \multicolumn{2}{c}{FMPE} 
            \\
    Trainable network   & \multicolumn{2}{c|}{CFM} 
                        & ViT+CFM 
                        & CNN+CFM 
                        & \multicolumn{2}{c}{CFM}
                        \\
    Epochs  & 337
            & 339  
            & 88  
            & 91
            & 714
            & 319
            \\
    Learning rate $\times10^{-4}$   & 5
                                    & 5
                                    & 5
                                    & 5
                                    & 5
                                    & 10
                                    \\
    Scheduler                 & \multicolumn{2}{c|}{Constant}
                              & \multicolumn{2}{c|}{Cosine Annealing}
                              & Constant 
                              & Cosine Annealing
                              \\
    Batch size  & 256
                & 256
                & 256
                & 256
                & 1024
                & 512
                \\
    Summary dimension   & 48+215
                        & 32+215
                        & 48+215
                        & 32+215
                        & 840
                        & 215 
                        \\
    Data    & \multicolumn{2}{c|}{SKA + Planck}
            & \multicolumn{2}{c|}{SKA + Planck}
            & SKA & Planck \\
    Training time  & 9m
                & 8m
                & 5h\,21m
                & 6h\,6m
                & 10m
                & 8m
                \\ 
    \bottomrule
  \end{tabular}
  \caption{Training hyperparameters across different stages, networks and data. Stage 1 - 3 are used when a summary network and 21\,cm data is used. The optional stages 4 and 5 are for joint inference. For 1dPS and Planck data only a single stage is required. A more detailed explanation can be found in Section~\ref{sec:impl}.}
  \label{tab:hyper_train}
\end{table}


\bibliographystyle{JHEP}
\bibliography{references}

\providecommand{\href}[2]{#2}\begingroup\raggedright\begin{thebibliography}{10}

\bibitem{Furlanetto:2006jb}
S.~Furlanetto, S.P.~Oh and F.~Briggs, \emph{{Cosmology at Low Frequencies: The 21 cm Transition and the High-Redshift Universe}}, \href{https://doi.org/10.1016/j.physrep.2006.08.002}{\emph{Phys. Rept.} {\bfseries 433} (2006) 181}.

\bibitem{sato_inflation}
K.~Sato, \emph{First-order phase transition of a vacuum and the expansion of the universe}, \href{https://doi.org/10.1093/mnras/195.3.467}{\emph{Monthly Notices of the Royal Astronomical Society} {\bfseries 195} (1981) 467}.

\bibitem{Starobinsky:1980te}
A.A.~Starobinsky, \emph{{A New Type of Isotropic Cosmological Models Without Singularity}}, \href{https://doi.org/10.1016/0370-2693(80)90670-X}{\emph{Phys. Lett. B} {\bfseries 91} (1980) 99}.

\bibitem{Starobinsky:1983zz}
A.A.~Starobinsky, \emph{{The Perturbation Spectrum Evolving from a Nonsingular Initially De-Sitter Cosmology and the Microwave Background Anisotropy}}, {\emph{Sov. Astron. Lett.} {\bfseries 9} (1983) 302}.

\bibitem{Planck:2018_inflation}
Y.~Akrami et~al., \emph{{Planck 2018 results. X. Constraints on inflation}}, \href{https://doi.org/10.1051/0004-6361/201833887}{\emph{Astron. Astrophys.} {\bfseries 641} (2020) A10}.

\bibitem{papamakarios2018fastepsilonfreeinferencesimulation}
G.~Papamakarios and I.~Murray, \emph{Fast $\epsilon$-free inference of simulation models with bayesian conditional density estimation},  2018.

\bibitem{Cranmer_2020}
K.~Cranmer, J.~Brehmer and G.~Louppe, \emph{The frontier of simulation-based inference}, \href{https://doi.org/10.1073/pnas.1912789117}{\emph{Proceedings of the National Academy of Sciences} {\bfseries 117} (2020) 30055–30062}.

\bibitem{neutsch:2022hmv}
S.~Neutsch, C.~Heneka and M.~Br{\"u}ggen, \emph{{Inferring astrophysics and dark matter properties from 21 cm tomography using deep learning}}, \href{https://doi.org/10.1093/mnras/stac218}{\emph{Mon. Not. Roy. Astron. Soc.} {\bfseries 511} (2022) 3446}.

\bibitem{Schosser:2024aic}
B.~Schosser, C.~Heneka and T.~Plehn, \emph{{Optimal, fast, and robust inference of reionization-era cosmology with the 21cmPIE-INN}}, \href{https://doi.org/10.21468/SciPostPhysCore.8.2.037}{\emph{SciPost Phys. Core} {\bfseries 8} (2025) 037}.

\bibitem{Ore:2024jim}
A.~Ore, C.~Heneka and T.~Plehn, \emph{{SKATR: A Self-Supervised Summary Transformer for SKA}}, \href{https://doi.org/10.21468/SciPostPhys.18.5.155}{\emph{SciPost Phys.} {\bfseries 18} (2025) 155}.

\bibitem{Planck:2019nip}
N.~Aghanim et~al., \emph{{Planck 2018 results. V. CMB power spectra and likelihoods}}, \href{https://doi.org/10.1051/0004-6361/201936386}{\emph{Astron. Astrophys.} {\bfseries 641} (2020) A5}.

\bibitem{Villaescusa:2022}
F.~{Villaescusa-Navarro}, J.~{Ding}, S.~{Genel}, S.~{Tonnesen}, V.~{La Torre}, D.N.~{Spergel} et~al., \emph{{Cosmology with One Galaxy?}}, \href{https://doi.org/10.3847/1538-4357/ac5d3f}{\emph{apj} {\bfseries 929} (2022) 132} [\href{https://arxiv.org/abs/2201.02202}{{\ttfamily 2201.02202}}].

\bibitem{Reza:2024djq}
M.~Reza, Y.~Zhang, C.~Avestruz, L.E.~Strigari, S.~Shevchuk and F.~Villaescusa-Navarro, \emph{{Constraining Cosmology with Simulation-based inference and Optical Galaxy Cluster Abundance}}, .

\bibitem{Zubeldia:2025qlt}
{\'I}.~Zubeldia, B.~Bolliet, A.~Challinor and W.~Handley, \emph{{Extracting cosmological information from the abundance of galaxy clusters with simulation-based inference}}, .

\bibitem{vonWietersheim-Kramsta:2024cks}
M.~von Wietersheim-Kramsta, K.~Lin, N.~Tessore, B.~Joachimi, A.~Loureiro, R.~Reischke et~al., \emph{{KiDS-SBI: Simulation-based inference analysis of KiDS-1000 cosmic shear}}, \href{https://doi.org/10.1051/0004-6361/202450487}{\emph{Astron. Astrophys.} {\bfseries 694} (2025) A223}.

\bibitem{Hothi:2023abe}
I.~Hothi, E.~Allys, B.~Semelin and F.~Boulanger, \emph{{Wavelet-based statistics for enhanced 21cm EoR parameter constraints}}, \href{https://doi.org/10.1051/0004-6361/202348444}{\emph{Astron. Astrophys.} {\bfseries 686} (2024) A212}.

\bibitem{Prelogovic:2023uww}
D.~Prelogovi{\'c} and A.~Mesinger, \emph{{Exploring the likelihood of the 21-cm power spectrum with simulation-based inference}}, \href{https://doi.org/10.1093/mnras/stad2027}{\emph{Mon. Not. Roy. Astron. Soc.} {\bfseries 524} (2023) 4239}.

\bibitem{Saxena:2023tue}
A.~Saxena, A.~Cole, S.~Gazagnes, P.D.~Meerburg, C.~Weniger and S.J.~Witte, \emph{{Constraining the X-ray heating and reionization using 21-cm power spectra with Marginal Neural Ratio Estimation}}, \href{https://doi.org/10.1093/mnras/stad2659}{\emph{Mon. Not. Roy. Astron. Soc.} {\bfseries 525} (2023) 6097}.

\bibitem{Zhao:2023uvf}
X.~Zhao, S.~Zuo and Y.~Mao, \emph{{3D ScatterNet: Inference from 21 cm Light-cones}},  in \emph{{40th International Conference on Machine Learning}}, 7, 2023.

\bibitem{Zhao:2023tep}
X.~Zhao, Y.~Mao, S.~Zuo and B.D.~Wandelt, \emph{{Simulation-based Inference of Reionization Parameters from 3D Tomographic 21 cm Light-cone Images. II. Application of Solid Harmonic Wavelet Scattering Transform}}, \href{https://doi.org/10.3847/1538-4357/ad5ff0}{\emph{Astrophys. J.} {\bfseries 973} (2024) 41}.

\bibitem{Zhao:2021ddh}
X.~Zhao, Y.~Mao, C.~Cheng and B.D.~Wandelt, \emph{{Simulation-based Inference of Reionization Parameters from 3D Tomographic 21 cm Light-cone Images}}, \href{https://doi.org/10.3847/1538-4357/ac457d}{\emph{Astrophys. J.} {\bfseries 926} (2022) 151}.

\bibitem{Bull:2014rha}
P.~Bull, P.G.~Ferreira, P.~Patel and M.G.~Santos, \emph{{Late-time cosmology with 21cm intensity mapping experiments}}, \href{https://doi.org/10.1088/0004-637X/803/1/21}{\emph{Astrophys. J.} {\bfseries 803} (2015) 21}.

\bibitem{Liu:2019srd}
A.~Liu et~al., \emph{{Cosmology with the Highly Redshifted 21cm Line}}, {\emph{Bull. Am. Astron. Soc.} {\bfseries 51} (2019) 63}.

\bibitem{Bacon2020SKA}
D.J.~Bacon, R.A.~Battye, P.~Bull, S.~Camera, P.G.~Ferreira, I.~Harrison et~al., \emph{Cosmology with phase 1 of the square kilometre array red book 2018: Technical specifications and performance forecasts}, \href{https://doi.org/10.1017/pasa.2019.51}{\emph{Publications of the Astronomical Society of Australia} {\bfseries 37} (2020) e007}.

\bibitem{Autieri:2025sxz}
G.~Autieri, M.~Berti, M.~Spinelli, B.S.~Haridasu and M.~Viel, \emph{{Weighing neutrinos with 21cm Intensity Mapping at the SKAO}}, .

\bibitem{Heneka:2018kgn}
C.~Heneka and L.~Amendola, \emph{{Constraining General Modifications of Gravity during Reionisation}},  in \emph{{53rd Rencontres de Moriond on Cosmology}}, pp.~207--210, 2018.

\bibitem{Liu:2019ygl}
X.-W.~Liu, C.~Heneka and L.~Amendola, \emph{{Constraining coupled quintessence with the 21cm signal}}, \href{https://doi.org/10.1088/1475-7516/2020/05/038}{\emph{JCAP} {\bfseries 05} (2020) 038}.

\bibitem{Liu:2015txa}
A.~Liu, J.R.~Pritchard, R.~Allison, A.R.~Parsons, U.~Seljak and B.D.~Sherwin, \emph{{Eliminating the optical depth nuisance from the CMB with 21 cm cosmology}}, \href{https://doi.org/10.1103/PhysRevD.93.043013}{\emph{Phys. Rev. D} {\bfseries 93} (2016) 043013}.

\bibitem{Modak:2022gol}
T.~Modak, L.~R{\"o}ver, B.M.~Sch{\"a}fer, B.~Schosser and T.~Plehn, \emph{{Cornering extended Starobinsky inflation with CMB and SKA}}, \href{https://doi.org/10.21468/SciPostPhys.15.2.047}{\emph{SciPost Phys.} {\bfseries 15} (2023) 047}.

\bibitem{Drees:2025ngb}
M.~Drees and Y.~Xu, \emph{{Refined predictions for Starobinsky inflation and post-inflationary constraints in light of ACT}}, \href{https://doi.org/10.1016/j.physletb.2025.139612}{\emph{Phys. Lett. B} {\bfseries 867} (2025) 139612}.

\bibitem{km3q-rm34}
D.S.~Zharov, O.O.~Sobol and S.I.~Vilchinskii, \emph{Act observations, reheating, and starobinsky and higgs inflation}, \href{https://doi.org/10.1103/km3q-rm34}{\emph{Phys. Rev. D} {\bfseries 112} (2025) 023544}.

\bibitem{Lehman:2024vyl}
K.~Lehman, S.~Krippendorf, J.~Weller and K.~Dolag, \emph{{Learning Optimal and Interpretable Summary Statistics of Galaxy Catalogs with SBI}}, .

\bibitem{Lanzieri:2024mvn}
D.~Lanzieri, J.~Zeghal, T.L.~Makinen, A.~Boucaud, J.-L.~Starck and F.~Lanusse, \emph{{Optimal neural summarization for full-field weak lensing cosmological implicit inference}}, \href{https://doi.org/10.1051/0004-6361/202451535}{\emph{Astron. Astrophys.} {\bfseries 697} (2025) A162}.

\bibitem{Makinen:2024xph}
T.L.~Makinen, C.~Sui, B.D.~Wandelt, N.~Porqueres and A.~Heavens, \emph{{Hybrid Summary Statistics}}, .

\bibitem{schiller:2025}
C.~{Heneka}, F.~{Nieser}, A.~{Ore}, T.~{Plehn} and D.~{Schiller}, \emph{{Large Language Models -- the Future of Fundamental Physics?}}, \href{https://doi.org/10.48550/arXiv.2506.14757}{\emph{arXiv e-prints} (2025) arXiv:2506.14757} [\href{https://arxiv.org/abs/2506.14757}{{\ttfamily 2506.14757}}].

\bibitem{Prelogovic:2024ips}
D.~Prelogovi{\'c} and A.~Mesinger, \emph{{How informative are summaries of the cosmic 21 cm signal?}}, \href{https://doi.org/10.1051/0004-6361/202449309}{\emph{Astron. Astrophys.} {\bfseries 688} (2024) A199}.

\bibitem{Diao:2024wrf}
K.~Diao, Z.~Chen, X.~Chen and Y.~Mao, \emph{{Reionization Parameter Inference from 3D Minkowski Functionals of the 21 cm Signals}}, \href{https://doi.org/10.3847/1538-4357/ad6c40}{\emph{Astrophys. J.} {\bfseries 974} (2024) 141}.

\bibitem{Wolz:2023gql}
K.~Wolz, N.~Krachmalnicoff and L.~Pagano, \emph{{Inference of the optical depth to reionization {\ensuremath{\tau}} from Planck CMB maps with convolutional neural networks}}, \href{https://doi.org/10.1051/0004-6361/202345982}{\emph{Astron. Astrophys.} {\bfseries 676} (2023) A30}.

\bibitem{FrancoAbellan:2025fkb}
G.~Franco~Abell{\'a}n, N.~Anau~Montel, O.~Savchenko and C.~Weniger, \emph{{How to embed any likelihood into SBI: Application to Planck + Stage IV galaxy surveys and Dynamical Dark Energy}}, .

\bibitem{kallosh2025presentstatusinflationarycosmology}
R.~Kallosh and A.~Linde, \emph{On the present status of inflationary cosmology},  2025.

\bibitem{DeFelice:2010aj}
A.~De~Felice and S.~Tsujikawa, \emph{{f(R) theories}}, \href{https://doi.org/10.12942/lrr-2010-3}{\emph{Living Rev. Rel.} {\bfseries 13} (2010) 3}.

\bibitem{Baumann:2009ds}
D.~Baumann, \emph{{Inflation}},  in \emph{{Theoretical Advanced Study Institute in Elementary Particle Physics}: {Physics of the Large and the Small}}, pp.~523--686, 2011, \href{https://doi.org/10.1142/9789814327183_0010}{DOI}.

\bibitem{Bassett:2005xm}
B.A.~Bassett, S.~Tsujikawa and D.~Wands, \emph{{Inflation dynamics and reheating}}, \href{https://doi.org/10.1103/RevModPhys.78.537}{\emph{Rev. Mod. Phys.} {\bfseries 78} (2006) 537}.

\bibitem{PhysRevLett.130.171403}
M.~Dax, S.R.~Green, J.~Gair, M.~P\"urrer, J.~Wildberger, J.H.~Macke et~al., \emph{Neural importance sampling for rapid and reliable gravitational-wave inference}, \href{https://doi.org/10.1103/PhysRevLett.130.171403}{\emph{Phys. Rev. Lett.} {\bfseries 130} (2023) 171403}.

\bibitem{lipman2023flowmatchinggenerativemodeling}
Y.~Lipman, R.T.Q.~Chen, H.~Ben-Hamu, M.~Nickel and M.~Le, \emph{Flow matching for generative modeling},  2023.

\bibitem{dax2023flowmatchingscalablesimulationbased}
M.~Dax, J.~Wildberger, S.~Buchholz, S.R.~Green, J.H.~Macke and B.~Schölkopf, \emph{Flow matching for scalable simulation-based inference},  2023.

\bibitem{NEURIPS2018_69386f6b}
R.T.Q.~Chen, Y.~Rubanova, J.~Bettencourt and D.K.~Duvenaud, \emph{Neural ordinary differential equations},  in \emph{Advances in Neural Information Processing Systems}, S.~Bengio, H.~Wallach, H.~Larochelle, K.~Grauman, N.~Cesa-Bianchi and R.~Garnett, eds., Curran Associates, Inc.

\bibitem{Planck:2018nkj}
{\scshape Planck} collaboration, \emph{{Planck 2018 results. I. Overview and the cosmological legacy of Planck}}, \href{https://doi.org/10.1051/0004-6361/201833880}{\emph{Astron. Astrophys.} {\bfseries 641} (2020) A1} [\href{https://arxiv.org/abs/1807.06205}{{\ttfamily 1807.06205}}].

\bibitem{Diego_Blas_2011}
D.~Blas, J.~Lesgourgues and T.~Tram, \emph{The cosmic linear anisotropy solving system (class). part ii: Approximation schemes}, \href{https://doi.org/10.1088/1475-7516/2011/07/034}{\emph{Journal of Cosmology and Astroparticle Physics} {\bfseries 2011} (2011) 034–034}.

\bibitem{Murray:2020trn}
S.G.~Murray, B.~Greig, A.~Mesinger, J.B.~Mu\~noz, Y.~Qin, J.~Park et~al., \emph{{21cmFAST v3: A Python-integrated C code for generating 3D realizations of the cosmic 21cm signal}}, \href{https://doi.org/10.21105/joss.02582}{\emph{J. Open Source Softw.} {\bfseries 5} (2020) 2582}.

\bibitem{Planck:2018vyg}
N.~Aghanim et~al., \emph{{Planck 2018 results. VI. Cosmological parameters}}, \href{https://doi.org/10.1051/0004-6361/201833910}{\emph{Astron. Astrophys.} {\bfseries 641} (2020) A6}.

\bibitem{Mao:2008ug}
Y.~Mao, M.~Tegmark, M.~McQuinn, M.~Zaldarriaga and O.~Zahn, \emph{{How accurately can 21 cm tomography constrain cosmology?}}, \href{https://doi.org/10.1103/PhysRevD.78.023529}{\emph{Phys. Rev. D} {\bfseries 78} (2008) 023529} [\href{https://arxiv.org/abs/0802.1710}{{\ttfamily 0802.1710}}].

\bibitem{Prince:2019hse}
H.~Prince and J.~Dunkley, \emph{{Data compression in cosmology: A compressed likelihood for Planck data}}, \href{https://doi.org/10.1103/PhysRevD.100.083502}{\emph{Phys. Rev. D} {\bfseries 100} (2019) 083502}.

\bibitem{21cmSense1}
J.C.~{Pober}, A.R.~{Parsons}, D.R.~{DeBoer}, P.~{McDonald}, M.~{McQuinn}, J.E.~{Aguirre} et~al., \emph{{The Baryon Acoustic Oscillation Broadband and Broad-beam Array: Design Overview and Sensitivity Forecasts}}, \href{https://doi.org/10.1088/0004-6256/145/3/65}{\emph{aj} {\bfseries 145} (2013) 65}.

\bibitem{21cmSense2}
J.C.~{Pober}, A.~{Liu}, J.S.~{Dillon}, J.E.~{Aguirre}, J.D.~{Bowman}, R.F.~{Bradley} et~al., \emph{{What Next-generation 21 cm Power Spectrum Measurements can Teach us About the Epoch of Reionization}}, \href{https://doi.org/10.1088/0004-637X/782/2/66}{\emph{apj} {\bfseries 782} (2014) 66}.

\bibitem{loshchilov2019decoupledweightdecayregularization}
I.~Loshchilov and F.~Hutter, \emph{Decoupled weight decay regularization},  2019.

\bibitem{brinckmann2018montepython3boostedmcmc}
T.~Brinckmann and J.~Lesgourgues, \emph{Montepython 3: boosted mcmc sampler and other features},  2018.

\bibitem{fisher1922foundations}
R.A.~Fisher, \emph{On the mathematical foundations of theoretical statistics}, \href{https://doi.org/10.1098/rsta.1922.0009}{\emph{Philosophical Transactions of the Royal Society of London. Series A, Containing Papers of a Mathematical or Physical Character} {\bfseries 222} (1922) 309}.

\bibitem{talts2020validatingbayesianinferencealgorithms}
S.~Talts, M.~Betancourt, D.~Simpson, A.~Vehtari and A.~Gelman, \emph{Validating bayesian inference algorithms with simulation-based calibration},  2020.

\bibitem{lemos2023samplingbasedaccuracytestingposterior}
P.~Lemos, A.~Coogan, Y.~Hezaveh and L.~Perreault-Levasseur, \emph{Sampling-based accuracy testing of posterior estimators for general inference},  2023.

\end{thebibliography}\endgroup
\end{document}